\documentclass[usenatbib]{mn2e}
\usepackage{epsfig}
\usepackage{amsmath,amssymb}
\usepackage{aas_macros}

\usepackage[dvips]{color}
\definecolor{rev}{rgb}{0.8,0.0,0.0}
\definecolor{rev2}{rgb}{0.0,0.0,0.8}
\definecolor{cut}{rgb}{1.0,0.547,0.0}

\def\vb#1{\mbox{\boldmath $#1$}}

\newcommand{\Caffau}{$\rm SDSS \ J102915+172927$}

\newcommand{\Aoki}{$\rm SDSS \ J001820.5-093939.2$}

\newcommand{\percc}{{\rm cm^{-3}}}
\newcommand{\um}{{\rm \mu m}}
\newcommand{\E}[1]{\times 10^{#1}}
\newcommand{\dex}{{\rm dex}}

\newcommand{\Enstatite}{{\rm MgSiO_3}}
\newcommand{\Forsterite}{{\rm Mg_2SiO_4}}

\newcommand{\Silica}{{\rm SiO_2}}
\newcommand{\Troilite}{{\rm FeS}}
\newcommand{\Alumina}{{\rm Al_2O_3}}
\newcommand{\Magnesia}{{\rm MgO}}

\newcommand{\nH}{n_{{\rm H}}}

\newcommand{\nHcen}{n_{\rm H,cen}}
\newcommand{\nHpeak}{n_{\rm H,peak}}

\newcommand{\namb}{n_{{\rm amb}}}

\newcommand{\Zsun}{{\rm Z_{\bigodot}}}
\newcommand{\Msun}{\rm M_{\bigodot}}

\newcommand{\Mpr}{M_{{\rm pr}}}

\newcommand{\fdepini}{f_{{\rm dep},0}}

\newcommand{\Mvir}{M_{\rm vir}}
\newcommand{\Rvir}{R_{\rm vir}}
\newcommand{\Vvir}{V_{\rm vir}}
\newcommand{\eturb}{\varepsilon _{\rm turb}}
\newcommand{\gammath}{\gamma _{\rm th}}

\defcitealias{Chiaki15}{C15}
\defcitealias{Hirano14}{H14}

\title[Low-metallicity gas collapse]
      {Gravitational collapse and the thermal evolution
        of low-metallicity gas clouds in the early Universe}

\author[G. Chiaki et al.]
{Gen Chiaki,$^{1, 2}$\thanks{E-mail: chiaki@center.konan-u.ac.jp}
Naoki Yoshida$^{1, 3}$ and
Shingo Hirano$^{1, 4}$
\\
$^{1}$Department of Physics, Graduate School of Science, University of Tokyo, 
7-3-1 Hongo, Bunkyo, Tokyo 113-0033, Japan \\
$^{2}$Department of Physics, Konan University,
8-9-1 Okamoto, Kobe, 658-0072, Japan \\
$^{3}$Kavli Institute for the Physics and Mathematics of the Universe (WPI), 
Todai Institutes for Advanced Study, \\
The University of Tokyo, Kashiwa, Chiba 277-8583, Japan \\
$^{4}$Department of Astronomy, The University of Texas, Austin, TX 78712, USA
}

\begin{document}

\date{}

\pagerange{\pageref{firstpage}--\pageref{lastpage}} \pubyear{2015}

\maketitle

\label{firstpage}

\begin{abstract}
We study gravitational collapse of low-metallicity gas clouds and the formation 
of protostars by three-dimensional hydrodynamic simulations. 
Grain growth, non-equilibrium chemistry, molecular cooling, and chemical heating
are solved in a self-consistent manner for the first time. 
We employ the realistic initial conditions for the abundances of metal and dust, and 
the dust size distribution obtained from recent Population III supernova calculations. 
We also introduce the state-of-the-art particle splitting method based on the Voronoi 
tessellation and achieve an extremely high mass resolution of $\sim 10^{-5} \ \Msun$ (10 Earth masses) 
in the central region. We follow the thermal evolution of several clouds with various metallicities. 
We show that the condition for cloud fragmentation depends not only on the gas metallicity 
but also on the collapse timescale. 
In many cases, the cloud fragmentation is prevented by the chemical heating owing to molecular 
hydrogen formation even though dust cooling becomes effective. 
Meanwhile, in several cases, efficient OH and H$_2$O cooling promotes the cloud elongation, and 
then cloud ``filamentation'' is driven by dust thermal emission as a precursor of eventual fragmentation. 
While the filament fragmentation is driven by rapid gas cooling with metallicity $\gtrsim 10^{-5} \ \Zsun$, 
fragmentation occurs in a different manner by the self-gravity of a circumstellar disk with metallicity $\lesssim 10^{-5} \ \Zsun$. 
We use a semi-analytic model to estimate the number fraction of the clouds which undergo 
the filament fragmentation to be 20--40\% with metallicity $10^{-5}$--$10^{-4} \ \Zsun$. 
Overall, our simulations show a viable formation path of the recently discovered 
Galactic low-mass stars with extremely small metallicities.
\end{abstract}



\section{INTRODUCTION}

The origin of Galactic extremely metal-poor (EMP) stars remains largely unknown. 
Their small metal content suggests that they were born very early,
while their small masses pose interesting questions on star formation
in a metal-poor environment.
For example, the star \Caffau \ with a mass $0.8 \ \Msun$
has an extremely low metallicity
of $4.5\E{-5} \ \Zsun$ \citep{Caffau11,Caffau12},
and thus is thought to be a second generation star;
it was probably born in a gas cloud that had been enriched with heavy elements
synthesized by the first generation of stars (Pop III stars).

Recent theoretical studies based on cosmological hydrodynamic simulations
suggest that the first generation stars formed from a metal-free, primordial
gas have a wide range of masses
\citep{Hirano14,Susa14}. 
Such a notion is supported by observations
of EMP stars whose peculiar abundance patterns are
consistent with the nucleosynthetic
models of massive primordial stars \citep{Caffau11,Keller14}.
Interestingly, the recently discovered
star \Aoki \ shows characteristic signatures of the metal yield
of a very massive primordial star,
suggesting a broad mass distribution
of Pop III stars \citep{Aoki14}.
In addition, it is reported that the fragmentation
of accretion disk in the primordial gas clouds leads the
solar- and subsolar-mass star formation when clumps are ejected
from the dense region by the $N$-body gravitational interactions \citep{Greif11,
Greif12,Clark11,StacyBromm14}.
If their mass is less than $\sim 1 \ \Msun$, these stars might be observed
in the local Universe \citep{Hartwig15, Ishiyama16}, and the slightly polluted
Pop III stars might be observed as metal-poor stars \citep{Komiya15}.

A trace amount of heavy elements in star-forming gas
plays an important role in the formation of low-mass stars with extremely
low metallicities \citep{Omukai00,BrommLoeb03,Schneider03}.
Efficient gas cooling from metal and dust can trigger the cloud fragmentation.
In general, gas cooling promotes cloud deformation 
into sheets or filaments, whereas heating processes stabilize the gas
against aspherical perturbations.
Linear analyses by \citet{Lai00} and \citet{HanawaMatsumoto00} 
demonstrate that a contracting cloud gets highly elongated
when the specific heat ratio 
$\gamma = {\rm d} \ln T / {\rm d} \ln \rho + 1$, an indicator
of the effective gas equation of state, is less than 1.1.
In a cooling gas with $\gamma < 1$, significantly elongated
filamentary structures tend to form.
When the density of the filament increases up
to a value where gas cooling becomes insufficient ($\gamma > 1$),
multiple cores are formed and begin contracting separately.
Therefore, the characteristic mass
of the fragments is set by the Jeans mass
at the density and temperature where the cooling becomes inefficient.

Analytic models and simulations suggest that intermediate-mass cores
($\sim 100 \ \Msun$) are formed
by transition line cooling of C {\sc i}, C {\sc ii}, and O {\sc i} 
\citep{BrommLoeb03,Frebel05,SantoroShull06,SafranekShrader14a}.
Low-mass stars might be formed by the dynamical ejection
with $N$-body interactions among these clumps \citep{Ji14}.
Dust thermal emission is another important process that determines
the thermal evolution of a low-metallicity gas.
A semi-analytic approach reveals that dust cooling becomes important at densities 
$\nH \gtrsim 10^{12} \ \percc$, where the corresponding Jeans mass is $\sim 0.1 \ \Msun$
\citep{Omukai00,Schneider03}.

Three-dimensional hydrodynamic simulations also have been performed 
to study the conditions for gas cloud fragmentation and for the formation
of low-mass stars. 
\citet{Jappsen07, Jappsen09a, Jappsen09b} investigate the process of low-metal gas collapse
with a wide range of parameter set.
They find that gas metallicity has little effect on cloud evolution and fragmentation 
even with metallicities around $10^{-3} \ \Zsun$ in the earlier stage of collapse ($\nH < 10^5 \ \percc$)
in fossile H {\sc ii} region created by a Pop III star.
Rather, the other parameter such as 
\citet{Dopcke11,Dopcke13} follow the dynamical evolution of a turbulent
gas with metallicities below $10^{-4} \ \Zsun$. They show that
several tens of sink particles are formed in the gas and that
the mass distribution of the sink particles transitions from flat to peaky distribution
with increasing metallicity.
\citet{SafranekShrader14a,SafranekShrader14b} perform simulations 
with a proper cosmological setup to investigate star formation in dense 
clumps in early galaxies with gas metallicities
of $10^{-4}$--$10^{-2} \ \Zsun$.
\citet{SafranekShrader14b} find that, with metallicity $10^{-2} \ \Zsun$, 
the gas fragments into sink particles,
whose mass distribution is consistent
with observationally derived stellar masses
in ultra faint dwarf spheroidal galaxies as observed by \citet{Geha13}.
More recently, \citet{SafranekShrader16} resume simulations with an updated
protostar model, and find that the mass distribution of sink particles resembles the
Salpeter mass function.
\citet{Smith15} study the effects of metal pollution and turbulence driven by
a Pop III supernova (SN) explosion on a neighboring halo.
The gas polluted to metallicity of $2\E{-5} \ \Zsun$ fragments
by dust thermal emission.
Their study explicitly includes
the process of the dispersion of metals and dust.

These previous studies assume the metal abundances,
grain condensation efficiency, 
and the dust size distribution to be those
in the Galactic interstellar medium
despite the fact that
the EMP stars observed so far show peculiar abundance patterns
\citep[also see the recently submitted paper of][]{Bovino16}.
For example, the abundance ratio of light elements,
such as carbon and oxygen relative to iron
is enhanced with respect to the solar value \citep{Suda08,Tominaga14}.
Because the efficiency of metal line cooling is determined by the abundance of
individual elements, it is important to adopt realistic values based
on, for example, the nucleosynthesis 
calculations of Pop III SNe.

The same is true for the condensation efficiency, or the mass fraction of heavy elements
condensed into dust grains. It is known that
a significant fraction of heavy elements are locked into dust grains in the local interstellar 
medium \citep{Pollack94}, but recent spectroscopic observations of
damped Lyman-$\alpha$ systems with metallicity $\sim 10^{-3} \ \Zsun$
suggest that the condensation efficiency is smaller at redshifts $z\sim 6$
\citep{Schady10,Zafar11}.
It is also known that grain growth,
i.e., accretion of the gas-phase metals onto dust grains, can enhance
the dust cooling efficiency in a collapsing cloud
even with an extremely low metallicity
to be close to that of the local interstellar
medium \citep{Nozawa12,Chiaki13}.

There are other important processes in low-metallicity gas that
have often been overlooked in previous studies. 
O-bearing molecules such as OH and H$_2$O act as important
coolants at intermediate densities $\nH \sim 10^4$--$10^8 \ \percc$.
Although OH and H$_2$O cooling is less effective in a gas
with a solar composition \citep{Jappsen07, Omukai00},
our previous work \citep[][hereafter \citetalias{Chiaki15}]{Chiaki15} reveals that 
the molecular cooling is efficient in some cases
when the realistic initial condition (composition) is adopted.
There is yet another important heating process.
When hydrogen molecules are formed, the binding energy is released to be converted
into the gas kinetic energy.
\citet{TsuribeOmukai08} argue that this gas heating can stabilize the gas to 
prevent fragmentation using simulations of a polytropic gas mimicking the 
thermal evolution of low-metallicity clouds.

Metal and dust properties in the early Universe
can be determined theoretically under
the assumption that the formation sites of low-metallicity stars
are in clouds partly polluted with metals and dust 
by Pop III SNe in the same or nearby halos \citep{Ritter12,Smith15}.
Thus, the abundances of heavy elements and dust species, and the dust size distribution
are expected to be set by nucleosynthetic process in SN explosion \citep{Umeda02}
and the dust formation and destruction history during the propagation of blast 
waves \citep{Todini01,Nozawa03,Schneider06}.

In this work, we perform hydrodynamic simulations
of collapsing gas clouds with various
metallicities $10^{-6}$--$10^{-3} \ \Zsun$.
We study in detail the critical conditions for gas fragmentation
and for the formation of low-mass protostars.
Our three-dimensional simulations follow, for the first time, 
all the necessary thermal processes including dust thermal emission,
gas heating by H$_2$ formation, and OH and H$_2$O cooling.
The metal abundances, initial grain condensation efficiency, 
and the dust size distribution are calculated from 
stellar evolution and SN explosion models of a Pop III star
with $30 \ \Msun$.
We calculate the chemical reactions and grain growth
in a direct, self-consistent manner in order to compute the cooling
and heating rates.
Recently, \citet{Hirano14} find that the different collapse timescales of the primordial 
star-forming cloud gives the wide mass distribution of Pop III stars, $\sim 10$--$1000 \ \Msun$.
In this paper, we employ four gas clouds with different collapse timescales to
examine the effect of the cloud variation on low-metallicity star formation.


\section{Numerical simulations}

\subsection{Chemistry and cooling}

We use the parallel $N$-body/smoothed particle hydrodynamics (SPH) 
code {\sc gadget}-2 \citep{Springel05}
with non-equilibrium chemistry and radiative cooling.
We solve chemical networks of 54 reactions for 27 gas-phase species:
H$^+$, $e^-$, H, H$^-$, H$_2$, D$^+$, D, HD,
C$^+$, C, CH, CH$_2$, CO$^+$, CO, CO$_2$, O$^+$, O, OH$^+$, OH, 
H$_2$O$^+$, H$_2$O, H$_3$O$^+$, O$_2^+$, O$_2$, Si, SiO, and SiO$_2$.
The chemical reaction rates are given in 
\citetalias{Chiaki15} for Si-bearing species, and
in \citet{Omukai10} for other species.
We solve the chemical reactions implicitly to obtain 
the abundances of the gas-phase species in each fluid element at each time step.
We then calculate the associated cooling and heating rates.

We implement radiative cooling by C$^+$, C, and O,
and by H$_2$ and HD molecules.
We calculate the level populations for each species
in a time-dependent manner.
Gas opacity is explicitly calculated and the cooling rate for each
emission line is reduced by a factor determined by the local velocity gradient 
in the optically thick regime (the so-called Sobolev approximation).
We calculate the velocity gradient in the three directions $x$, $y$, and $z$ as in \citet{Hirano13}.
We also include metal molecular line cooling, using the formulation and the cooling tables presented 
by \citet{Neufeld93} and \citet{Neufeld95} for H$_2$O and \citet{Omukai10} for CO and OH.
We obtain the cooling rate $\Lambda _i (m)$ of molecules $m= \rm CO$, OH, and H$_2$O with 
the parameter $\tilde N (m) = n(m) / |dv_i/dr_i|$ corresponding to optical depth under the large velocity gradient
(LVG) approximation
in the three directions $i=x$, $y$, and $z$, and take the average as 
$\Lambda (m) = \left[ \Lambda _x (m) + \Lambda _y (m) + \Lambda _z (m) \right] / 3$.

We calculate the optical depth for the continuum emission 
as $\tau _{\rm con} = (\kappa _{\rm g} \rho _{\rm g} + \alpha _{\rm d}) \lambda _{\rm J}$, where 
$\kappa _{\rm g}$ and $\rho _{\rm g}$ are the absorption cross-section of the primordial 
gas \citep{Mayer05} and mass density of the gas component, respectively.
The absorption coefficient $\alpha _{\rm d}$ for grains is calculated for all grain size and species as
Equation (8) of \citetalias{Chiaki15}.
To save the computational cost, we use the one-zone model with the shielding length
$\lambda _{\rm J} = (\pi c_{\rm s}^2 / G \rho)^{1/2}$.
The emission rate of each transition line is further reduced by a factor of $e^{-\tau _{\rm con}}$.
We consider the continuum emission by collision-induced emission (CIE) of hydrogen
molecules \citep{Yoshida06} and by dust grains.
These continuum emotion rates are multiplied by escape fraction 
$\beta _{\rm con} = \min \{ 1, \tau _{\rm con} ^{-2} \}$
\citep{Omukai00}.

Hydrogen molecules are formed via endothermic reactions
with $E = 4.48$ eV per formed molecule.
The three-body reactions particularly enhance the temperature at 
densities $10^8$--$10^{11} \ \percc$.
\citet{TsuribeOmukai08} find that the gas heating resulting from the
rapid hydrogen molecular formation mitigates the cloud deformation 
and makes the cloud core rounder.
We consider the heating mechanism according to the formulation of
\citet{HollenbachMcKee79} and \citet{Omukai00}.

\subsection{Grain growth}

The dust species considered here are
metallic silicon (Si), metallic iron (Fe), forsterite ($\Forsterite$), 
enstatite ($\Enstatite$), amorphous carbon (C),
silica ($\Silica$), magnesia ($\Magnesia$), troilite ($\Troilite$),
and alumina ($\Alumina$), which are major species condensed in
unmixed SN ejecta \citep{Nozawa03, Nozawa07}.
We follow the evolution of the size distributions
and abundances of the grain species by
essentially treating the grain growth as chemical reactions.
We consider 13 reactions as in \citetalias{Chiaki15}.
We integrate the growth rate \citepalias[Equation 9 in][]{Chiaki15} 
to derive the grain radius and condensation efficiency
at each time for each fluid element.
We then calculate
the reaction rate of hydrogen molecular formation on grain surfaces, 
radiative cooling efficiency, and 
continuum opacity
for every grain species and radius.

\subsection{Particle splitting}
We need to follow the gas cloud collapse to very high densities.
The Jeans length of the central, densest part decreases eventually
down to $\sim 0.1$ AU when a protostellar core is formed.
To save the computational cost, we use the particle splitting technique,
when the resolution is about to violate the Jeans criterion
\citep{Truelove97, Truelove98}. Dense gas particles are replaced with
less massive daughter particles.
In \citet{ChiakiYoshida15}, we present a novel method in which
the daughter particles are distributed
based on the Voronoi diagram tessellated by parent particles.
With this method, the density structures of the cloud is well preserved.
In the entire course of our simulations, a Jeans mass is required
to be resolved by more than $1000 M_{\rm min}$,
where $M_{\rm min} = N_{\rm ngb} m_{\rm p}$ is the minimal resolvable
mass of the parent SPH particles with mass $m_{\rm p}$,
and $N_{\rm ngb}$ is the number of the neighbor particles.
We set $N_{\rm ngb} = 64 \pm 8$, and thus the Jeans mass is always resolved by
$\sim 10^5$ particles.

\begin{table*}
\begin{minipage}{17cm}
\caption{Properties of gas clouds}
\label{tab:cloud}
\begin{tabular}{cccccccccc}
\hline
Cloud & $z_{\rm form}$ & $\Mvir^{\rm dm}$ & $\Mvir^{\rm ba}$ & $\Rvir$ & $M_{\rm PopIII}$ &
$\alpha$ & $\beta$ & $\lambda$ & $\eturb$ \\
\hline \hline
UNI  & ---        & $0.0\E{0}$ & $2.2\E{6}$ & 551 &      --- & 0.51 & 0.0069 & 0.015 & 0.350 \\
MH1 & 20.46 & $1.4\E{5}$ & $2.0\E{4}$ & 26.5 & 283.9 & 0.46 & 0.0049 & 0.011 & 0.086 \\
MH2 & 16.20 & $1.4\E{5}$ & $3.7\E{4}$ & 46.8 & 751.3 & 0.48 & 0.0320 & 0.050 & 0.069 \\
MH3 & 15.15 & $8.1\E{4}$ & $1.6\E{4}$ & 37.5 &   60.5 & 0.99 & 0.0325 & 0.063 & 0.067 \\
\hline
\end{tabular}
\medskip \\
Note --- $z_{\rm form}$ is formation redshift of the minihalos.
$\Mvir^{\rm dm}$ [$\Msun$] and $\Mvir^{\rm ba}$ [$\Msun$] are mass of 
the dark matter and baryon, respectively, within the Virial radius 
$\Rvir$ [pc] within which the average density is 200 times cosmological average.
$\alpha = \bar e \Rvir / G\Mvir$ and $\beta = 5 E_{\rm rot} \Rvir / 3G \Mvir ^2$ are the ratios of thermal and 
rotational energy to gravitational energy, where $\bar e$ is the mass-weighted average of specific energy
over SPH particles within $\Rvir$, $E_{\rm rot} $ is the sum of the rotational energy
$m | {\vb r} \times {\vb v} |^2 / 2r^2$ of an SPH particle with mass $m$, coordinate ${\vb r}$,
and velocity ${\vb v}$, and $\Mvir = \Mvir^{\rm dm}+\Mvir^{\rm ba}$ is the total mass within radius $\Rvir$.
$\lambda = \bar j/\sqrt{2}\Vvir \Rvir$ is the spin parameter \citep{Bullock01}, where 
$\bar j$ is the mass-weighted average of specific angular momentum $j=|{\vb r}\times {\vb v}|$
and $\Vvir = \sqrt{G \Mvir / \Rvir}$ is the circular velocity at the Virial radius.
$\eturb = \bar {\vb v}_{\rm turb}^2/(\bar {\vb v}^2+\bar c_{\rm s} ^{2})$ is the ratio of the turbulent energy to the 
internal energy, where $\bar {\vb v}_{\rm turb}$ is the root-mean-square of velocity as used by \citet{MacLow99},
and $\bar {\vb v}$ and $\bar c_{\rm s}$ are the mass-weighted averages of kinetic and sound velocities, respectively.
These values are measured when the peak density reaches $\nHpeak = 10^{-3} \ \percc$
for metallicity $Z=10^{-4} \ \Zsun$.
\end{minipage}
\end{table*}

\begin{table*}
\begin{minipage}{17cm}
\caption{Properties of heavy elements and dust grains}
\label{tab:metal}
(a) Abundance of heavy elements \\
\begin{tabular}{cccccccc}
\hline
$X$ & C & O & Mg & Al & Si & S & Fe  \\
\hline \hline
$A_X$ & $ 1.08\E{-4}$ & $ 1.19\E{-3}$ & $ 4.19\E{-5}$ & $ 8.29\E{-7}$ & $ 6.67\E{-5}$ & $ 3.01\E{-5}$ & $ 6.76\E{-6}$ \\
$[X/{\rm Fe}]$ &  $ 0.22$ & $     1.01$ & $     0.77$ & $     0.14$ & $     0.99$ & $     1.01$ & $     0.00$ \\
\hline
\end{tabular}
\medskip \\
Note --- Abundances of heavy elements released by a Pop III SN
with progenitor mass $30 \ \Msun$ in an ambient gas with density
$\namb = 1 \ \percc$ \citep{Nozawa07}. 
We show the number abundances relative to hydrogen nuclei with 
metallicity $Z=Z_{\bigodot}$.
The values are reduced by the factor $Z/Z_{\bigodot}$ for metallicity $Z$.
In the second row, the relative abundance 
$[X/{\rm Fe}] = \log (A_X/A_{\rm Fe}) - \log (A_{X, \bigodot } / A_{{\rm Fe}, \bigodot })$
is shown.
The solar abundance $A_{X, \bigodot } $ is taken from \citet{Caffau11a}. \\ \\
(b) Composition and characteristic size of grain species \\
\begin{tabular}{crrrrrrrrrr}
\hline
$i$ & Silicon & Iron & Forsterite & Enstatite & Carbon & Silica & Magnesia & Troilite & Alumina & Total \\   
\hline \hline
$f_{{\rm dep,}i,0} \ [\times 10^{-3}]$ & $  29.14$ & $   1.93$ & $   0.77$ & $ <0.01$ & $   0.62$ & $   5.27$ & $   1.34$ & $   0.45$ & $ <0.01$ & $  39.53$ \\
$r_{i,0}^{\rm grow} \ [\times 10^{-2} \ \um]$ & $  38.67$ & $  28.30$ & $   1.20$ & $   1.38$ & $   1.97$ & $   5.76$ & $   5.31$ & $   3.53$ & $   0.11$ & \\
\hline
\end{tabular}
\medskip \\
Note --- Initial depletion factor $f_{{\rm dep,}i,0}$ relative to total mass of metal and dust and 
the characteristic radius $r_{i,0}^{\rm grow} = \langle r^3 \rangle _i / \langle r^2 \rangle _i$
of grain species $i$.
\end{minipage}
\end{table*}

\subsection{Later accretion phase}
\label{sec:AccPhase}
As the density in the central core increases, the dynamical time 
decreases and the necessary integration time step gets progressively shorter.
Sink particle techniques are often employed to save computational time,
where the gas inside a pre-determined
accretion radius is replaced with a sink particle 
\citep{Dopcke13,SafranekShrader14b,Smith15}.
The evolution of a circumstellar disk, through which
the gas is accreted, is not followed accurately with sink particle techniques.
It is also known that the separation of fragments depends
on the accretion radius 
\citep{MachidaDoi13,Greif12}.
Instead of employing a sink particle technique, we resort to
following the gas dynamics in and around the proto-stellar core
by setting the specific heat ratio to be $\gamma =1.4$
after the density of fluid elements exceeds $\nH = 10^{16} \ \percc$, where
the gas becomes optically thick even in a primordial gas.
Although we are not able to follow the thermal evolution accurately beyond
$\nH = 10^{16} \ \percc$, the treatment allows us
to examine gas fragmentation owing to dust thermal emission,
which is expected to be effective at $\nH \gtrsim 10^{12}$--$10^{15} \ \percc$.

\subsection{Initial conditions}

\subsubsection{Cloud models}
We use two different sets of initial conditions; one spherical cloud
as a controlled simulation, and three clouds hosted by
small dark matter halos
selected from cosmological simulations.

\vspace{5mm}
\noindent{\bf Uniform density cloud}

The spherical cloud has a uniform density 
$n_{\rm H,ini}=0.1 \ \percc$ and temperature $T_{\rm ini} = 300 \ {\rm K}$.
The cloud radius and mass are $R_{\rm ini}=551$ pc and $M_{\rm ini}=2.3\E{6} \ \Msun$,
corresponding to the half Jeans length and the Jeans mass, respectively, for the uniform cloud with 
density $n_{\rm H,ini}/f_{\rm enh}$, where $f_{\rm enh} = 1.8$ is the enhancement factor
\citep{MatsumotoHanawa03}.
We first put SPH particles in a box with 
length $2R_{\rm ini}$ on a side with a random seed.
Since the amplitude of density perturbation resulting from Poisson noise 
is too large, we perform the preliminary calculation without gravity
to relax the perturbation.
The calculation is performed with the isothermal equation of state 
in a periodic boundary box until the root-mean-square
of particle density becomes 0.1 of the mean density 
\citep[the same level as the simulations of][]{Machida15}.
Then we cut out the gas sphere with radius $R_{\rm ini}$.

We impose solid rotation such that
the rotation energy relative to the gravitational energy 
$\beta _{\rm ini}= \Omega _{\rm ini}^2 R_{\rm ini}^3 / 3GM_{\rm ini}$ is $10^{-3}$, 
corresponding to the angular velocity $\Omega _{\rm ini}=1.4\E{-17} \ {\rm s^{-1}}$.
The initial number fractions of deuterium and helium relative to hydrogen nuclei
are $5.3\E{-5}$ and $0.079$, respectively.
The number fractions of electron and hydrogen molecules relative to hydrogen nuclei are 
$y(e)=10^{-4}$ and $y({\rm H_2})=10^{-6}$, respectively.
Both deuterium and helium are initially assumed to be in neutral atoms.
The spherical cloud calculations are dubbed ``UNI'' hereafter.
The cosmic microwave background (CMB) radiation contribute to 
the gas heating by exciting the level populations and giving the energy with grains.
For UNI, we assume the CMB temperature $T_{\rm CMB} = 45$ K,
corresponding to the redshift $z\simeq 15$.

\vspace{5mm}
\noindent{\bf Minihalos}

We also perform collapse simulations for gas clouds hosted by 
minihalos chosen from the cosmological simulations of \citet{Hirano14} \citepalias[hereafter][]{Hirano14}.
From the large simulation box of $\sim 1 $ Mpc on a side,
we cut out the spherical region including dark matter particles
centered at a minihalo with radii $\sim 1$ kpc 
so that sound waves cannot cross the region during the cloud collapse. 
The abundances of primordial species are inherited
from the cosmological simulations.
The CMB temperature is calculated from
$T_{\rm CMB} = 2.73(1+z_{\rm form})$ K
at formation redshift $z_{\rm form}$ of the halos.

We select three minihalos dubbed ``MH1'', ``MH2'', and ``MH3'',
which cover the different types of the early clouds.
Their properties are presented in Table \ref{tab:cloud}.
The sixth column shows the final stellar mass $M_{\rm PopIII}$ derived by the collapse
simulations without metal performed by \citetalias{Hirano14}.
For MH1, MH2, and MH3, $M_{\rm PopIII} = 283.9$, $751.3$, and $60.5 \ \Msun$, respectively.
The difference of stellar mass is affected by the different
thermal evolution from clouds to clouds
as discussed in Appendix C of \citetalias{Hirano14} and below.

\vspace{5mm}
\noindent{\bf Cloud properties}

Various parameters can characterize gas clouds.
In Table \ref{tab:cloud}, we show the mass of baryon 
$\Mvir ^{\rm ba}$ and dark matter $\Mvir ^{\rm dm}$
within the Virial radius $\Rvir$ within which the average density is 200 times larger
than the cosmological average density and the over-dense region turns to collapse
against the cosmic expansion \citep{Peebles80}.
These parameters characterize the depth of potential well $G \Mvir / \Rvir$ created by
the halos, where $\Mvir = \Mvir ^{\rm ba} + \Mvir ^{\rm dm}$.
The collapse timescale of cloud is expected to become longer with shallower potential.
As \citetalias{Hirano14} predict, the collapse time will affect the thermal evolution
of clouds.

The ratios $\alpha$ and $\beta$ of thermal and rotational energy
to the gravitational energy could also have effects on cloud fragmentation
as discussed by \citet{TsuribeInutsuka99a,TsuribeInutsuka99b}.
For UNI, MH1, and MH2, $\alpha \simeq 0.5$ and small $\beta$ ($\beta < 0.1$), which
predicts that the clouds are marginally unstable.
For MH3, large $\alpha$ ($ \simeq 1$) would lead the cloud to collapse monolithically.
However, we should note that the prediction is based on the semi-analytical model
where gas evolves isothermally \citep[see Figure 8 in][]{TsuribeInutsuka99b}.
The low-metal clouds no longer evolve isothermally.
Another parameter which characterizes the spin is $\lambda$, which is defined as
the ratio between angular momentum of clouds and that of gravitationally bound object.
\citet{Jappsen07} find that the parameter have little effects on the cloud evolution
in the earlier stage of collapse ($\nHcen < 500 \ \percc$) when $\lambda \lesssim 0.05$.
We aim to see how the cloud rotation has effect on the later stage of collapse.

In Table \ref{tab:cloud}, we also present the parameter of turbulent motion $\eturb$
as the ratio of turbulent energy to the internal energy.
For minihalos (MH1, MH2, and MH3), the clouds are less turbulent ($\eturb=0.067$--$0.086$)
while the random density perturbation is converted to the large turbulent motion for UNI 
when the peak density is $10^3 \ \percc$ ($\eturb = 0.350$).
The turbulence might affect the fragmentation properties of the clouds \citep{Dopcke13, Smith15}.

\begin{figure}
\includegraphics[width=8.5cm]{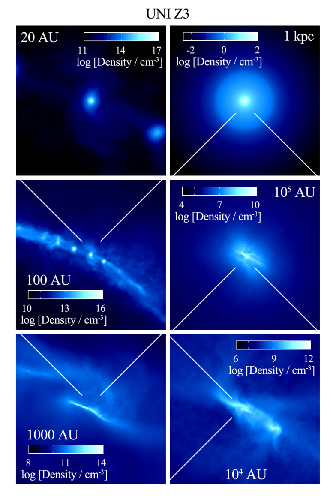}
\caption{
Zoom-in distribution of the density-weighted density projection 
centered at the most dense region of
the cloud UNI Z3 at 36.4 years after the first protostar formation.
The side length of the panels becomes shorter clockwise from top-right (1 kpc) to
top-left (20 AU).
This cloud fragments into two clumps owing to the OH molecular
cooling at the length scale $\sim 1000$ AU (bottom-right panel).
The inner part of the left clump further fragments into three protostellar cores and several diffuse
blobs by dust thermal emission at the length scale $\sim 10 $ AU (middle-left).}
\label{fig:snapshot_ps_UN300_Z-3}
\end{figure}

\begin{figure*}
\includegraphics[width=18cm]{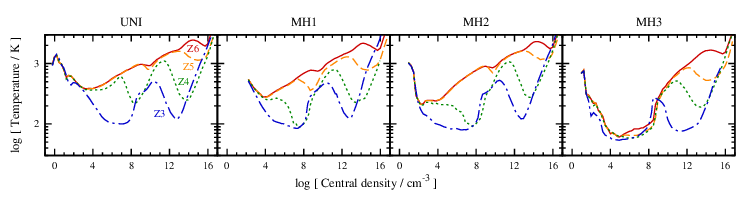}
\caption{
Thermal evolution of the cloud cores 
with metallicities $10^{-6}$ (Z6; red solid), 
$10^{-5}$ (Z5; orange dashed), $10^{-4}$ (Z4; green dotted), 
and $10^{-3} \ \Zsun$ (Z3; blue dot-dashed)
for clouds UNI, MH1, MH2, and MH3 from left to right.}
\label{fig:nT_f3}
\end{figure*}

\begin{figure}
\includegraphics[width=9cm]{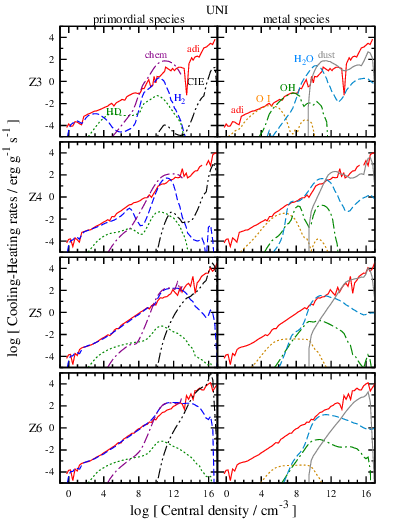}
\caption{
Thermal evolution of cooling/heating rates
for Z6, Z5, Z4, and Z3 from top to bottom for cloud UNI.
Left and right panels show the cooling efficiencies of the primordial and metal
species, respectively.
``adi'' depicts the adiabatic compressional heating rate, and
``chem'' depicts the absolute value of net heating/cooling efficiency owing to
H$_2$ formation/dissociation.}
\label{fig:nL_UN300_f3}
\end{figure}

\subsubsection{Metal and dust properties}

In our simulations, metals and dust are uniformly added to
the simulation gas particles.
The metal abundance, dust condensation efficiency, 
and dust size distribution are taken from a model of 
nucleosynthesis and grain formation/destruction in a Pop III SN.
For the first time, we adopt such the realistic initial conditions
to the three-dimensional simulations of the Pop II star formation,
while previous works often assume the solar abundance pattern.
We employ a Pop III SN model taken from \citet{Nozawa07}.
We here make use of the model with the progenitor
mass $30 \ \Msun$ and the ambient gas density $\namb = 1 \ \percc$ 
as a characteristic model.

The progenitor mass determines the abundance of heavy elements.
The mass spectrum of Pop III star has the bimodal distribution
as predicted by \citetalias{Hirano14} and \citet{Hirano15}.
In their simulations, the first peak lies around $30 \ \Msun$.
\citet{Susa14} also predict the peaky distribution with characteristic 
mass $20$--$30 \ \Msun$.
In our simulation, employed $\Mpr$ is limited to $30 \ \Msun$.
We discuss the fragmentation property for clouds enriched by progenitors with
various masses in Section \ref{sec:discussion}.

The dust grains are formed in expanding SN ejecta, and destroyed
by the reverse shocks.
The strength of the destruction depends on the ambient gas density $\namb$.
Just before the explosion, $\namb$ is determined by the interaction between gas and ultraviolet
photons emitted from a main-sequence (MS) Pop III star.
The H {\sc ii} region generally has the nearly uniform density by the strong radiation pressure
with $\namb = 0.1$--$1 \ \percc$ \citep{Kitayama04, Whalen04}.
Here, we conservatively set the ambient gas density as $1 \ \percc$.

We briefly summarize the difference between the Pop III
SN model and the present-day model.
First, the number fractions of the light elements relative to iron 
are larger in the latter case (second row of Table \ref{tab:metal}a).
In particular, the Pop III model predicts a large fraction of
$\rm [O/Fe] \simeq 1$.
We thus expect that the radiative cooling by OH and H$_2$O molecules is more efficient
than in the case of solar abundance.
Second, the total condensation efficiency of metal is only 4\%, which is much
smaller than $\sim 50$\% in the present-day.
Although the grain growth can enhance the final dust-to-gas mass ratio,
the efficiency of H$_2$ formation on grains and dust cooling rate
are still smaller than in the present-day case.

The Pop III SN model gives us the relative abundances of metal and dust.
Their absolute values are determined for a given metallicity, or the total mass fraction
of heavy element to the gas.
We examine four cases with metallicities
$10^{-6}$, $10^{-5}$, $10^{-4}$, and $10^{-3}$ times the solar value $Z_{\bigodot} = 0.02$.
Hereafter we call the models as Z6, Z5, Z4, and Z3, respectively.
We assume that initially all of C nuclei are in the form of C$^+$,
and O and Si are in neutral atoms.
The initial mass fraction $f_{{\rm dep},i,0}$ of grain species $i$ relative to metal and
the characteristic radius of grains \citepalias[see][]{Chiaki15} are 
shown in Table \ref{tab:metal} (b).


\begin{figure*}
\includegraphics[width=15cm]{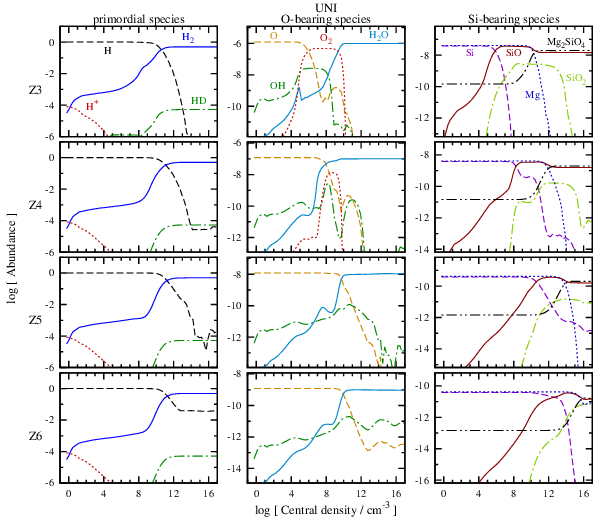}
\caption{
Chemical abundance of primordial species (left column),
O-bearing species (middle), and Si-bearing species (right)
for Z6, Z5, Z4, and Z3 from top to bottom for UNI.
}
\label{fig:ny_UN300_f3}
\end{figure*}

\section{Chemo-thermal evolution of spherical clouds}

Figure \ref{fig:snapshot_ps_UN300_Z-3} shows the zoom-in image of our simulated cloud
UNI Z3.
Particle splitting allows us to follow
the nonlinear gravitational evoltution from a diffuse interstellar
cloud to protostellar core(s) over 20 orders of magnitude in density.
The mass resolution $M_{\rm min}$ is initially $320 \ \Msun$ and eventually reaches
$3.2\E{-5} \ \Msun$ (10 Earth masses), corresponding to $\lesssim 0.06$ AU,
in the central densest region with $\gtrsim 10^{16} \ \percc$
after splitting particles seven times.
We in this section see the temperature evolution, comparing to the cooling/heating
efficiencies for UNI cloud.
Figure \ref{fig:nT_f3} shows the temperature evolution as a function of
the central density with metallicities $10^{-6}$--$10^{-3} \ \Zsun$ of 
clouds UNI, MH1, MH2, and MH3.\footnote{Hereafter, we measure the quantities
such as the temperature, density, chemical abundance, and cooling rate in the central region
by mass-weighting over the gas particles
with densities $\nH > \nHpeak/3$, 
where $\nHpeak$ is the largest density in each snapshot.}
Figures \ref{fig:nL_UN300_f3} and \ref{fig:ny_UN300_f3} show the cooling/heating efficiencies
and the abundances of gas- and solid-phase species 
as a function of the central density
for UNI Z3--6.

Adiabatic compressional heating is dominant for all the metallicity 
cases at densities $\lesssim 1 \ \percc$.
After that, molecular hydrogen cooling becomes important 
until the density reaches the critical value for H$_2$ molecules, $\nH \sim 10^3 \ \percc$,
where the level populations thermalize and approach their LTE values.
For Z4--6, the temperature increases only slowly by the balance between the gas compressional
heating and H$_2$ line cooling.
For Z3, the fine-structure line cooling by O {\sc i} becomes efficient.
Thereafter, OH molecules form via the reaction
\[ \rm O + H \to OH + \gamma , \ \ \ \ \ \ (Reaction \ 30) \]
and then OH cooling becomes dominant at $\nH \sim 10^4$--$10^8 \ \percc$ for Z3.
The gas fragments into two clumps by efficient OH cooling for Z3
as shown in Figure \ref{fig:snapshot_ps_UN300_Z-3}.

\begin{figure*}
\includegraphics[width=15cm]{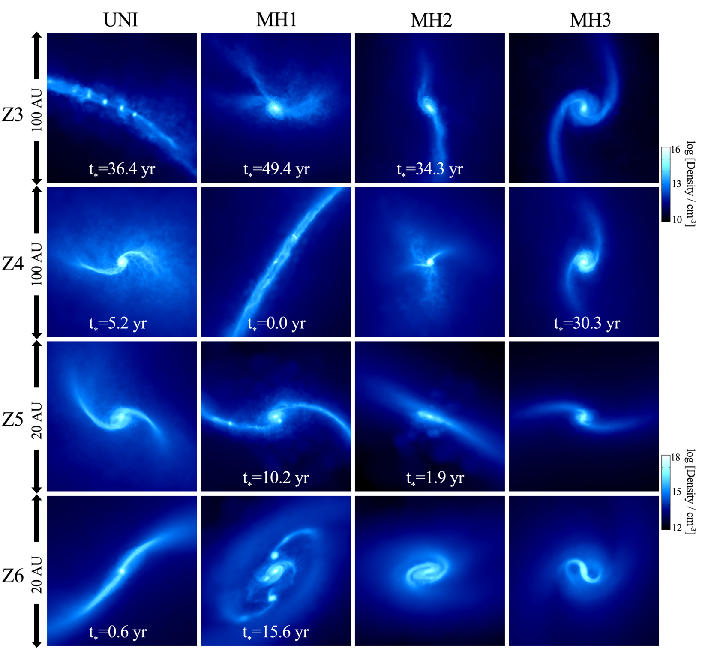}
\caption{
Density-weighted density projection of the clouds UNI, MH1, MH2, and MH3 from left to right
with metallicities $10^{-6}$, $10^{-5}$, $10^{-4}$, and $10^{-3} \ \Zsun$ from bottom to top
at the time where the simulations are terminated.
The plotted region is 100 AU (upper 8 panels) and 20 AU (lower 8 panels) on a side.
The color contour depicts the density $\nH = 10^{10}$--$10^{16} \ \percc$ (upper 8 panels)
or $10^{12}$--$10^{18} \ \percc$ (lower 8 panels) from black to white.
These snapshots are output at the time $t_*$ from the first protostar is formed
as written in the bottom of each pane.
In the models without the time, the central blob does not satisfy our criterion of a protostar 
(see footnote \ref{foot:Protostar}) until the end of the simulation.}
\label{fig:snapshot_all}
\end{figure*}

\begin{figure*}
\includegraphics[width=18cm]{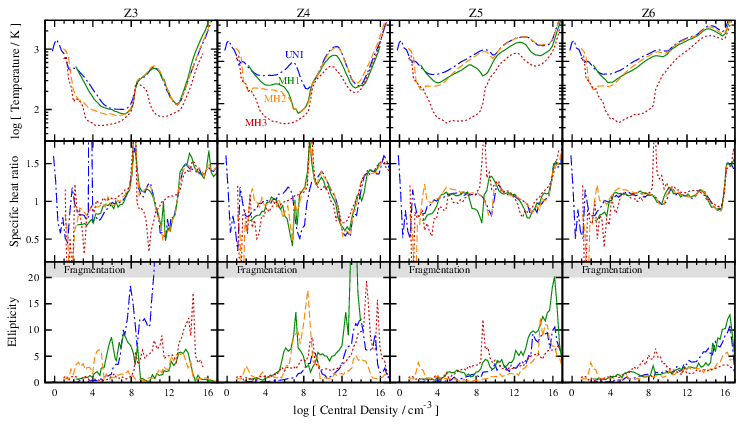}
\caption{
Temporal evolution of the temperature (top), specific heat ratio (middle) 
and ellipticity ${\cal E}$ (bottom) as a function 
of the central density of cloud cores UNI (blue dot-dashed), 
MH1 (green solid), MH2 (orange dashed), and MH3 (red dotted) 
for Z3--6 from left to right.
We measure the specific heat ratio as 
$\gamma = (\Delta T /T)/ ( \Delta \rho / \rho) + 1$,
where $\Delta x $ is the difference of the quantity $x$ between a snapshot
and its previous snapshot.
We arrange the output times such that the maximum density $\nHpeak$ increases 0.25 dex
between the two adjacent outputs.
The grey shaded region in the bottom panels indicates the ellipticities ${\cal E}>20$ (see text) for which the cloud
breaks up into multiple clumps.}
\label{fig:nTe_3d}
\end{figure*}

At the same time, hydrogen molecules are formed on grain 
surfaces.
The binding energy is released and converted into the gas thermal energy.
Although the heating rate of grain-surface reaction dominates over
that of the other reacitons, 
it never exceeds the cooling efficiency by rotational transition line cooling of OH molecules.
Only by the grain-surface reaction,
cloud cores do not become fully molecular until the density reaches $\sim 10^8 \ \percc$.
The molecular fraction reaches $3\E{-3}$.
For Z4--6, the molecules are formed only via H$^-$ processes as
\[ \rm H + e \to H^- + \gamma, \ \ \ \ \ \ (Reaction \ 2) \]
\[ \rm H^- + H \to H_2 + e, \ \ \ \ \ \ (Reaction \ 3) \]
and molecular fraction becomes $2\E{-4}$.
At density $\sim 10^8 \ \percc$, 
the three-body H$_2$ formation reactions
\[ \rm H + H + H \to H_2 + H, \ \ \ \ \ \ (Reaction \ 5) \]
\[ \rm H + H + H_2 \to H_2 + H_2  \ \ \ \ \ \ (Reaction \ 6) \]
operate, and the clouds get fully molecular.
At this moment, the exothermic reactions raise the gas temperature rapidly.
Then the specific heat ratio temporarily exceeds the critical value of 4/3
for gravitational collapse.
Then a pressure-supported core surrounded by accretion shock is formed
for Z3.
The velocity jump is seen at the rapid heating works.

The gas temperature continues to increase to $\sim 10^{11} \ \percc$ 
while the H$_2$O molecules act as a main coolant
The H$_2$O molecules are formed via the reactions
\[ \rm OH + H_2 \to H_2O + H . \ \ \ \ \ (Reaction \ 24) \]
For Z5, H$_2$O cooling is a dominant cooling process in a narrow
range of density around $10^{10} \ \percc$
as shown by the cyan curve in Figure \ref{fig:nL_UN300_f3}.

Dust cooling becomes efficient at $10^{11}$--$10^{13} \ \percc$ for Z3,
and at slightly larger densities $10^{12}$--$10^{14} \ \percc$ for Z4.
The grain growth enhances the efficiency of the dust thermal emission
for Z3--5.
Forsterite grains grow rapidly via the reaction
\[ \rm 2 Mg (g) + SiO (g) + 3 H_2O (g) \to Mg_2SiO_4 (s) + 3H_2 (g), \]
where (g) and (s) denote the species in the gas- and solid-phases, respectively.
The magnesium atoms are eventually exhausted 
(see blue dotted curve in Figure \ref{fig:ny_UN300_f3}).
For Z6, although there is little amount of dust,
the temperature drops at $10^{14} \ \percc$ by collision-induced
continuum cooling (black dot-dot-dashed curve in Figure \ref{fig:nL_UN300_f3}).

Dust cooling becomes ineffective
when thermal coupling between the gas and grains is established.
The cloud core becomes optically thick soon after this phase,
at densities $1\E{13}$, $3\E{14}$, $3\E{15}$, and $1\E{16} \ \percc$ 
for Z3, Z4, Z5, and Z6, respectively.
The gas pressure rapidly increases in the region where radiative cooling
is inefficient due to the large opacity,
and then a hydrostatic core bounded by accretion shock is formed.
The second velocity jump appears in the velocity profile
as well as at the density where H$_2$ formation heating proceeds.
Just inside the shock, the gas disk is formed, being supported by the rotation.
As the temperature further increases toward the center, 
the central part becomes spherical because of the strong pressure support.
Here, let us define a pressure-supported ``protostar'' as the region 
bounded by a sufficiently spherical iso-density surface.\footnote{In practice, 
we calculate iso-density contours by
dividing the gas into 100 density bins equally separated
with a logarithmic scale from $\log (\nHpeak ) - 3$ to $\log (\nHpeak )$.
Fitting the distribution of the gas particles in each density bin by
an ellipsoid with the major- and minor-axes $a$ and $b$, 
we obtain the ellipticity ${\cal E}=a/b-1$ of each iso-density surface.
We then define the protostellar surface as the least dense (most distant from the center)
surface that satisfies ${\cal E} < {\cal E}_*$ with the critical ellipticity ${\cal E} _* = 0.3$.
The protostellar mass is only 20\% smaller with a smaller threshold of
${\cal E}_* = 0.2$.\label{foot:Protostar}}
The cloud fragmentation again occurs for Z3 around the time of the first protostar formation
as shown in Figure \ref{fig:snapshot_ps_UN300_Z-3}.

\section{Cloud fragmentation}

\subsection{Thermal evolution and fragmentation of cloud}
\label{sec:therm}

\subsubsection{Overview of the global feature}
In order to see whether the gas clouds fragment or not, we follow
the evolution over several tens years after the first protostar appears.
Figure \ref{fig:snapshot_all} shows the snapshots output at the time $t_*$
after the first protostar formation.
Contrary to the popular notion that cloud fragmentation conditions
are largely determined by the gas metallicity,
the fragmentation properties are different even 
with the same metallicity (see panels in each row for a given metallicity).
For example, for MH3 Z4, fragmentation does not occur even 30 years after
the first protostar formation,
while multiple clumps are already formed before the first protostar is formed for MH2 Z4.
Clearly, whether a cloud fragments or not is not determined
solely by the gas metallicity.

In other words, the thermal evolution, 
which critically affects the fragmentation property,
is not uniquely determined by the gas metallicity.
The top panels of Figure \ref{fig:nTe_3d} shows the temperature evolutions
as a function of the central density of four clouds for Z5 and Z4.
The evolutionary tracks on the $\nH$-$T$ plane vary significantly even
with a fixed metallicity.
The variation would generate the different fragmentation property as discussed
in more detail later.
We begin with discussing the physical processes that drive the variation of the thermal evolution
in Section \ref{sec:ColTime}.
Then we discuss the effect of the thermal evolution on the cloud fragmentation
in Section \ref{sec:CoolHeat}.
Finally, in Section \ref{sec:FragCri}, we derive criteria for cloud fragmentation.

\subsubsection{Filament fragmentation vs. disk fragmentation}
We note an important point that
the cloud fragmentation is observed also at a very 
small metallicity of $10^{-6} \ \Zsun$,
where dust cooling is not efficient.
For MH1 Z6, the perturbations grow in the spiral arms in a circumstellar 
disk with a size of $\simeq 10$ AU 
around the central most massive core.
Such a structure is reported also by \citet{Clark11}
and \citet{Greif12} in the case without metal or dust.
Whereas, the simulations with intermediate metallcities $10^{-5}$--$10^{-3} \ \Zsun$
show a different type of fragmentation.
For UNI Z3, MH1 Z4, and MH1 Z5,
the ``peapod''-like structure, where several
protostellar cores are hosted in a long filament extending over
$\sim 20$--100 AU, appears.
Such a peculiar structure is formed typically when dust cooling operates, 
as we discuss later.

Hereafter, we distinguish the former and latter types of
fragmentation as {\it disk fragmentation}
and {\it filament fragmentation}, respectively.
Since the disk fragmentation has been discussed in 
previous numerical studies \citep{Clark11,Greif12}, 
we in this paper focus on the filament fragmentation, and
revisit the disk fragmentation briefly in Section \ref{sec:DiskFrag}.

\subsection{Collapse timescale and the thermal evolution}
\label{sec:ColTime}

As described in Section 4.1.1, the thermal evolution has an effect
on the cloud fragmentation property even with a fixed metallicity.
We find that the different thermal evolution among clouds can be explained
by the variation of the collapse timescale.
In a slowly collapsing cloud,
the adiabatic compressional heating rate per unit volume 
\begin{equation}
\Gamma _{\rm adi} = -\rho p \frac{{\rm d}}{{\rm d}t} \left( \frac{1}{\rho} \right) = \frac{p}{t_{\rm col}}
\label{eq:Gadi}
\end{equation}
is correspondingly small. Here, $t_{\rm col}$ is the collapse timescale written as
\begin{equation}
t_{\rm col} = \frac{\rho }{ {\rm d}\rho /{\rm d}t }.
\end{equation} 
Therefore, the evolutionary track on the $\nH$-$T$ plane tends to be
located in a low-temperature region.

We measure in our simulations that the time interval over which the peak density increases from
$n_{\rm H, peak} = 10^{2} \ \percc$ to $10^{16} \ \percc$ is
3, 4, 11, and 19 Myr for UNI, MH1, MH2, and MH3 Z4, respectively.
The cloud UNI contracts most rapidly while
the cloud MH3 contracts most slowly among the four clouds.
The evolutionary tracks in Figure \ref{fig:nTe_3d} roughly
trace the trend;
the temperature of UNI remains high at all densities because of its slow
rotation and hence the most rapid collapse.
Contrastingly, the temperature of MH3 remains low because 
the large centrifugal force and the flat distribution of the dark matter 
slow the gas infall.
We will discuss the origin of the different 
collapsing time in Section \ref{sec:discussion}.

\begin{figure}
\includegraphics[width=8.5cm]{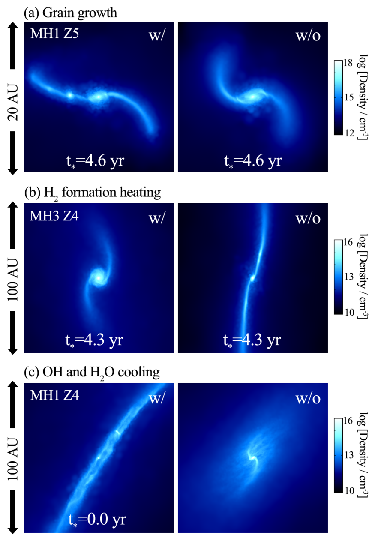}
\caption{
We compare the cloud fragmentation in the regular simulations (left) and controlled simulations (right)
at the same epoch.
All the relevant chemo-thermal processes are considered in the former simulations,
while, in the latter simulations,
(a) grain growth, 
(b) H$_2$ formation heating, and 
(c) OH and H$_2$O cooling
are not included.
}
\label{fig:snapshot_cool}
\end{figure}

\subsection{Important thermal processes}
\label{sec:CoolHeat}

Figure \ref{fig:nTe_3d} shows that
different cooling and heating processes become important
and characterize the various evolutionary tracks.
For example, MH1 and MH3 Z5 enter the regime where dust 
cooling is efficient at $\nH \sim 10^{12}$--$10^{14} \ \percc$ while
UNI and MH2 do not.
In addition to dust cooling, two other cooling and heating processes
responsible to the fragmentation can be identified:
H$_2$ formation heating and OH and H$_2$O cooling.
We discuss the processes one-by-one in the following.

\subsubsection{Dust cooling}
Dust cooling is a crucial process to lower the gas temperature,
to promote cloud elongation, and 
to induce the formation of low-mass fragments
at high densities $10^{12}$--$10^{15} \ \percc$,
where the Jeans mass is small $\sim 0.1 \ \Msun$.
Figure \ref{fig:nTe_3d} clearly shows the effect of dust cooling
on the filament formation for UNI Z3 and MH1 Z4 and Z5.
The bottom panels present the cloud ellipticity ${\cal E} = a/b - 1$,
where $a$ and $b$ are the major- and minor-axes of the cloud.
The gas becomes unstable owing to the dust cooling at
densities $\nH \sim 10^{12}$--$10^{14} \ \percc$,
and bar-mode perturbations grow.
A filamentary structure is formed as clearly seen in Figure \ref{fig:snapshot_all}.
When the ellipticity becomes above 20--30, which is consistent with the critical value
defined by \citet{TsuribeOmukai06},
the dense filament quickly fragments to yield several protostars.

For UNI and MH2 Z5,
even though the temperature decreases slightly by dust cooling
at $\nH\sim 10^{12}$--$10^{14} \ \percc$,
it is insufficient to enhance the ellipticity.
The cloud ellipticity increases to only at most 11 and 12 for UNI and MH2, respectively.
Figure \ref{fig:snapshot_all} shows that short filaments are formed in these
clouds, but without fragmenting to multiple clumps.

We emphasize that the growth of dust grains further enhances the
cooling rate and thus promotes the gas elongation and fragmentation.
In the Pop III SN dust model adopted here, $\Forsterite $ grains
grow until gas-phase Mg atoms are exhausted
at $\sim 10^{14} \ \percc$.
Figure \ref{fig:snapshot_cool} (a) compares our simulations with (left) and without (right)
grain growth for MH1 Z5 at the same time $t_* = 4.6$ yr after the first protostar formation.
While several protostellar cores are born from a filament in the
case with grain growth,
only a single protostar is formed at the center of a slightly more diffuse filament
otherwise.

\subsubsection{H$_2$ formation heating}

In many cases, dust cooling becomes efficient at high densities
but the gas cloud does not fragment. We find that chemical heating
effectively makes the cloud stable against deformation. 
For example, in MH3 Z5,
the rapid gas heating by the formation of H$_2$ molecules via three-body reactions
becomes efficient at $\nH \sim 10^9$--$10^{12} \ \percc$.
The temperature rapidly increases from 130 K at $\nH=5\E{8} \ \percc$
to 700 K at $2\E{11} \ \percc$,
causing the specific heat ratio to be 1.7 on average, which is sufficient
to stabilize the gas against elongation.
Then the gas ellipticity rapidly decreases from 12 to 2 
as shown in the bottom panel of Figure \ref{fig:nTe_3d}.

The timescale for which the ellipticity increase from such a small value to the critical
ellipticity is longer than the dynamical timescale of the cloud.
In the clouds UNI, MH1, and MH2 with metallicities $10^{-4}$--$10^{-3} \ \Zsun$ and in the cloud MH3
with $10^{-6}$--$10^{-3} \ \Zsun$, the H$_2$ formation heating significantly suppresses
the gas elongation before dust cooling 
becomes effective to promote the gas elongation.

\citet{TsuribeOmukai08} show that a gas cloud 
does not fragment with the metallicity range of $10^{-5}$--$10^{-4} \ \Zsun$,
by an order of magnitude smaller than in our simulations.
In their present-day dust model predicts 30 times larger efficiency of the
H$_2$ molecular formation on grains than in our Pop III SN dust model
(see Table \ref{tab:metal} b).
Hydrogen atoms are exhausted by the formation process before the molecular
formation via three-body reactions at $\nH \sim 10^8 \ \percc$.
In their model, H$_2$ formation heating therefore does not become efficient and
can not stabilize the gas.
Meanwhile, in our simulations even with $10^{-4}$--$10^{-3} \ \Zsun$,
three-body reactions and the associated gas heating make the cloud
rounder.

In order to see the effect more clearly,
we perform a controlled simulation without the heating process.
Figure \ref{fig:snapshot_cool} (b) shows the density distributions in the central 100 AU
region of the cloud MH3 Z4 with (left) and without (right) H$_2$ formation heating.
In the former case, 
the cloud core ellipticity ${\cal E}$ remains around 3
because of the effective chemical
heating at $\sim 10^8 \ \percc$.
In the latter case ${\cal E} $ increases
up to 15 by OH cooling and then further increases to $\sim 30$
by dust cooling without any stabilization.
Density perturbations in the filament grow and yield several protostellar cores
in a thin filament.

\begin{figure}
\includegraphics[width=9cm]{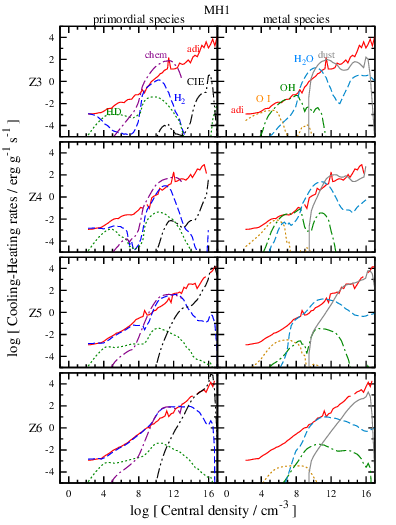}
\caption{
Same as Figure \ref{fig:nL_UN300_f3} but for
cloud MH1.}
\label{fig:nL_H0611_f3}
\end{figure}

\begin{figure*}
\includegraphics[width=15cm]{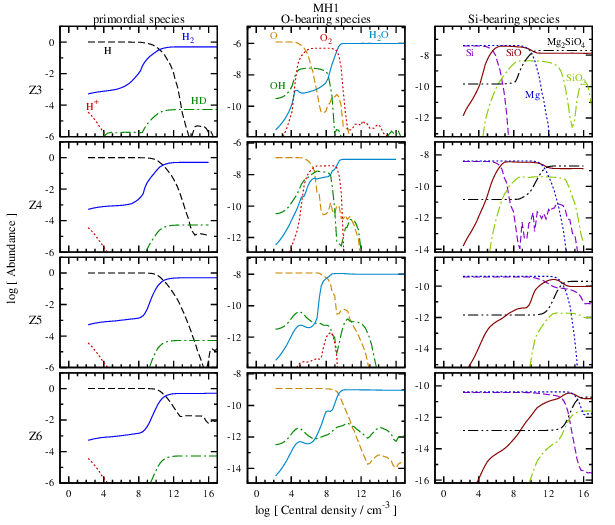}
\caption{
Same as Figure \ref{fig:ny_UN300_f3} but for
cloud MH1.}
\label{fig:ny_H0611_f3}
\end{figure*}

\subsubsection{OH and H$_2$O cooling}
\label{sec:OH_H2O}

In the cloud MH1 Z4, although H$_2$ formation heating is effective
(see the green solid curve in Figure \ref{fig:nTe_3d}),
a thin filament is formed and quickly fragments to a few clumps.
We find that radiative cooling owing to the transition lines of OH and H$_2$O molecules
is important in this case.
Figures \ref{fig:nL_H0611_f3} and \ref{fig:ny_H0611_f3} show the contribution of various heating and
cooling processes as a function of density for the cloud.
At $\nH = 10^6$--$10^8 \ \percc$, a large fraction of O atoms are converted
into OH molecules (green dot-dashed curve),
and OH cooling exceeds the adiabatic compressional heating rate (red solid).
Even though H$_2$ formation heating is effective at $\nH \sim 10^9$--$10^{11} \ \percc$
and reduces the cloud ellipticity,
${\cal E}$ evaluated at $10^{11} \ \percc$
is the largest among the four clouds for Z4 thanks to efficient OH
and H$_2$O cooling.
The cloud is then sufficiently elongated to trigger fragmentation.
The ellipticity increases up to 33 in the end (see Figure \ref{fig:nTe_3d}).
Although H$_2$O cooling is indeed effective for MH2 Z4 at $10^6$--$10^8 \ \percc$ 
(orange dashed curve in Figure \ref{fig:nTe_3d}),
it is not sufficient to compensate the stabilization effect by H$_2$ formation.

Again, in order to see clearly the effect of the OH and H$_2$O cooling,
we perform a controlled simulation without OH and H$_2$O 
cooling for MH1 Z4. The result is shown in
Figure \ref{fig:snapshot_cool} (c).
Without the molecular cooling, even at densities where dust cooling is efficient,
a single protostar is formed.
Clearly, the efficient OH and H$_2$O cooling drives
the formation of a long filament and subsequent fragmentation.
This is for the first time demonstrated by our simulations which explicitly
include the metal molecular cooling and 
the Pop III SN model with the large oxygen excess.

\begin{figure*}
\includegraphics[width=18cm]{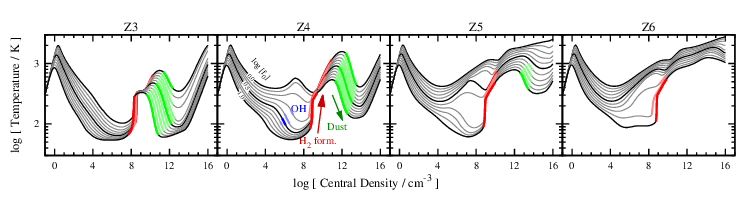}
\caption{
Temperature as a function of the density in a cloud core obtained by the semi-analytic chemo-thermal
evolution model of gas clouds
for Z3--6 from left to right
with $f_0$ (defined as Equation \ref{eq:f0}) from 1 to 10 every 0.5 dex (thick curves)
and every 0.1 dex (thin curves).
We also draw thick and colored segments in the regimes
where $\gamma > 1.1$ by H$_2$ formation heating (red), 
where $\gamma < 0.7$ by OH and H$_2$O cooling (blue), and 
where $\gamma < 0.8$ by dust emission (green)
on each evolutionary track.
}
\label{fig:nT_Z-04_0d}
\end{figure*}

\subsection{Criteria for first low-mass star formation}
\label{sec:FragCri}

From the discussion in Sections \ref{sec:ColTime}--\ref{sec:CoolHeat},
we can consider that the three thermal processes, dust cooling, H$_2$ formation heating, and
OH and H$_2$O cooling are important to determine the fragmentation
properties of the clouds.
Therefore, the criteria for gas fragmentation can be translated
into the condition where these processes become effective or not 
for a given combination of the metallicity $Z$ and the collapse timescale $t_{\rm col}$.
We define the fragmentation condition as
\begin{itemize}
\item[({\it i})] dust cooling is efficient at high densities $\sim 10^{12} \ \percc$ 
to induce eventual elongation and fragmentation of clouds, and
\item[({\it ii}-a)] H$_2$ formation heating is not efficient at intermediate
density $\sim 10^9$--$10^{11} \ \percc$, or 
\item[({\it ii}-b)] OH or H$_2$O cooling is efficient to trigger cloud elongation
at density $\sim 10^6$--$10^8 \ \percc$
even though H$_2$ formation heating later stabilizes the gas.
\end{itemize}

There remains a complexity that the collapse timescale itself varies
when the cloud collapse proceeds faster/slower
owing to gas cooling/heating.
We thus introduce a parameter $f_0$ as an indicator of the collapse 
timescale that characterizes each cloud.
The parameter satisfies the equation as
\begin{equation}
t_{\rm col} = f_0 t_{\rm col}^{\rm s} (\gamma ).
\label{eq:f0}
\end{equation}
The factor $t_{\rm col}^{\rm s} (\gamma )$ accounts for the
dependence on the specific heat ratio $\gamma$, the indicator of the thermal evolution.
At $\gamma > 0$, gas clouds undergo inhomogeneous collapse,
and the pressure gradient force delays collapse with respect to the
collapse time $t_{\rm col}^{\rm s} (0)$ in the limit of $\gamma \to 0$ as 
\begin{equation}
t_{\rm col}^{\rm s} (\gamma ) = \frac{1}{\sqrt{1-f_p (\gamma )}} t_{\rm col}^{\rm s} (0),
\end{equation}
where $f_p (\gamma )$ is the ratio of pressure gradient force to gravitational force \citep{Larson69}.
In this limit,
the cloud dynamics becomes practically identical with the dust collapse, 
in which the pressure is uniformly zero.
The collapse time asymptotically approaches 
$t_{\rm col}^{\rm s} (0) = (2/3\pi )t_{\rm ff} = 0.21t_{\rm ff}$, where
$t_{\rm ff} = ( 3\pi / 32 G\rho) ^{1/2}$ is the free-fall time at mass density $\rho$ \citep{Penston69}.
The ratio $f_p (\gamma )$ can be derived from the self-similar solution of spherical
clouds with polytropic equation of state.
\citet{Omukai05} present the fitting formula as
\begin{eqnarray}
f_p (\gamma )=  \begin{cases}
    0, & (\gamma < 0.83) \\
    0.6+2.5(\gamma -1) -6.0 (\gamma -1)^2, & (0.83<\gamma <1) \\
    1.0+0.2(\gamma - 4/3) -2.9 (\gamma - 4/3)^2. & (\gamma > 1)
  \end{cases}
\end{eqnarray}
Hereafter, we utilize the parameter $f_0$ to represent the bulk rate of the cloud collapse.

\subsubsection{One-zone semi-analytic model}

It is costly to run additional three-dimensional
simulations which cover a large parameter regions of $Z$ and $f_0$.
We thus resort to utilizing a simpler approach focusing
on the chemo-thermal evolution of a cloud.
To this aim, the semi-analytic one-zone model of \citetalias{Chiaki15} is suitable.
We modify the code such that  
the density in the cloud center increases according to 
\begin{equation}
\frac{{\rm d}\rho }{ {\rm d}t } = \frac{\rho }{ t_{\rm col} } = \frac{ \rho }{ f_0 t_{\rm col}^{\rm s} (\gamma )}
\end{equation}
with various $f_0$.\footnote{\citetalias{Chiaki15}
investigate the cloud evolution only in the case with $f_0 = 3\pi /2 = 4.7$ 
\citep[also see][]{Larson69, Omukai10}.}
With $\gamma \to 4/3$, the collapse time $t_{\rm col}^{\rm s} (\gamma )$ diverges.
Our simulations show that the clouds continues to collapse by accreting ambient gas
even with $\gamma > 4/3$.
To mimic this, we set the upper limit of $f_p (\gamma )$ as 0.95.
For given density and temperature, our one-zone code solves
exactly the same chemical reactions including grain growth
and gas heating/cooling processes as in our three-dimensional simulations.
The CMB temperature is set to be 50 K (redshift $\sim 20$).
We perform the calculations with varying $f_0=1$--10 every 0.05 dex and 
$Z=10^{-6}$--$10^{-3} \ \Zsun$ every 0.1 dex to cover the parameter
range of early clouds.

Figure \ref{fig:nT_Z-04_0d} shows the temperature evolution of the cloud center
with various values $f_0$ from 1 to 10 every 0.1 dex.
The gas temperature monotonically decreases with increasing $f_0$ at each density.
We find that the evolution track for our simulated clouds (Figure \ref{fig:nTe_3d})
follows closely one of the black curves in Figure \ref{fig:nT_Z-04_0d}
with a corresponding $f_0$ shown in Table \ref{tab:f0}.
We indicate by thick colored segments the regimes where we can identify 
important thermal processes that determine the cloud elongation and 
fragmentation.
Extending the approach of \citet{Schneider10}, we from our simulations define
the threshold value of specific heat ratio $\gammath$ below/above which
cooling/heating makes cloud elongated/round.

The gas cloud can fragment if the cloud ellipticity
is sufficiently large in advance to the rapid gas cooling.
In our simulations, $\gamma$ decline
down only to 0.84 and 0.85 for UNI and MH2 Z5, respectively,
while down to 0.78 for MH1 Z5
(see middle panels of Figure \ref{fig:nTe_3d}).
The critical $\gammath ^{\rm dust} $ for dust-induced fragmentation can be defined
as 0.8.
The green segments in Figure \ref{fig:nT_Z-04_0d} represent the regime where dust cooling is
efficient to reduce $\gamma $ down to 0.8.

The condition is not met if the trajectory passes through the regime 
colored in red, where significant H$_2$ formation heating processes.
Figure \ref{fig:nTe_3d} shows that the $\gamma$
increases larger than 1.1.
Although $\gamma$ exceed 1.1 for UNI Z3, 
the strong turbulence ($\eturb = 0.350$) as well as significant dust cooling
efficiency might induce fragmentation.
Let us define the threshold $\gammath ^{\rm H_2}$ as 1.1 for H$_2$ formation heating to halt
fragmentation.
In Figure \ref{fig:nT_Z-04_0d}, the red segments are plotted in the regime where $\gamma$ exceeds 1.1 due to
hydrogen molecular formation.

In the limited cases with metallicity $10^{-4} \ \Zsun$,
we find that the OH cooling, which is efficient in
the blue-colored regions at $\nH \sim 10^7 \ \percc$,
can prevent the gas cloud from recovering a round shape.
Although $\gamma$ reaches down to 0.41 at density 
$\nH = 4\E{6} \ \percc$ in our simulation (for MH1 Z4), 
it never decreases down to 0.7 in our one-zone calculations.
It would be the limitation of the one-zone model.
Since the threshold $\gammath ^{\rm OH}$ for OH cooling has an uncertainty,
we search $\gammath ^{\rm OH}$ which reproduce
the simulation results.
In Figure \ref{fig:nTe_3d}, we show the area with $\gammath ^{\rm OH} = 0.7$ as blue-colored
segments.
As the simulations with metallicity $10^{-4} \ \Zsun$ indicate, $\gamma$ becomes
below $\gammath^{\rm OH}$ with a limited range of $f_0$.
The cloud passes the regime where OH cooling is efficient
in the range $f_0 \simeq 2.5$--4.5.

\begin{table}
\caption{Collapse timescale of clouds $f_0$.}
\label{tab:f0}
\begin{tabular}{ccccc}
\hline
$Z \ [\Zsun ]$ & UNI & MH1 & MH2 & MH3  \\
\hline \hline
$10^{-3}$   & $ 1.1 \pm  0.4$ & $ 1.7 \pm  0.6$ & $ 2.4 \pm  2.0$ & $ 3.6 \pm  2.1$\\
$10^{-4}$   & $ 1.1 \pm  0.4$ & $ 2.2 \pm  0.9$ & $ 2.4 \pm  1.6$ & $ 5.4 \pm  3.4$\\
$10^{-5}$   & $ 1.0 \pm  0.3$ & $ 2.1 \pm  0.6$ & $ 2.0 \pm  1.8$ & $ 6.2 \pm  3.2$\\
$10^{-6}$   & $ 1.0 \pm  0.2$ & $ 2.0 \pm  0.5$ & $ 2.1 \pm  1.8$ & $ 6.4 \pm  3.4$\\
\hline
\end{tabular}
\medskip \\
Note --- To compute $f_0$ for the simulated clouds,
we measure the collapse time as $t_{\rm col} = \rho / (\Delta \rho / \Delta t)$
at each output time with $10^2 < \nHpeak <10^8 \ \percc$.
We then take the arithmetic mean and the standard deviation of $f_0$.
\end{table}

\begin{figure}
\includegraphics[width=8.5cm]{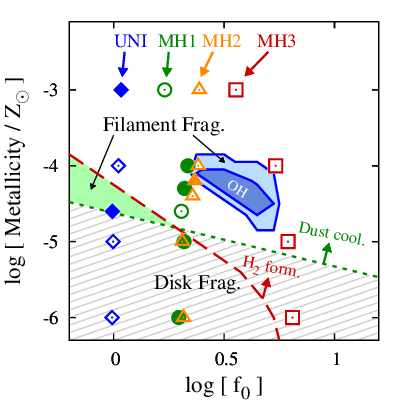}
\caption{
We plot the regions of the collapse timescale $f_0$ and the metallicity $Z$ favourable
for the filament fragmentation (green- and blue- shaded regions) and
for the disk fragmentation (grey-hatched region).
Dust cooling is efficient in the region above the green dashed lines (labeled ``Dust cool.'').
H$_2$ formation heating is efficient in the region above the red dotted lines (``H$_2$ form.'').
OH and H$_2$O cooling is efficient in the region surrounded by blue solid line
(``OH'').
We draw the lines for the threshold specific heat ratios $\gammath ^{\rm OH} = 0.65$ (blue)
and $0.70$ (cyan) above which OH cooling is regarded to promote sufficiently (see text).
We define these lines by linear interpolation of the results of the semi-analytic calculations 
with varying $f_0 = 1$--$10$ every 0.05 dex and $Z/ Z_{\bigodot} = 10^{-6}$--$10^{-3}$ every 0.1 dex.
We also over-plot $f_0$ and $Z$ of the simulated clouds 
UNI (blue diamonds),
MH1 (green circles),
MH2 (orange triangles), and
MH3 (red squares).
Open and close symbols indicate the models that end with and without
fragments, respectively.
}
\label{fig:ox_fZ}
\end{figure}

We present the results of all our model calculations
in Figure \ref{fig:ox_fZ}.
The green- and blue-shaded regions
indicate the regime where fragmentation criteria are satisfied.
There, we also mark the results of our three-dimensional simulations
(Table \ref{tab:f0}).
The models that result in fragmentation is indicated by filled symbols,
and otherwise by open symbols.
The parameter regions for the fragmentation defined by the semi-analytic model
are qualitatively consistent with the results of the simulations with $10^{-5}$--$10^{-4} \ \Zsun$.
The result for UNI Z3 can not be reproduced, but this may reflect 
simply the limitation of our one-zone model to represent the non-linear evolution of 
hydrostatic core formed by H$_2$ formation heating.

In Figure \ref{fig:ox_fZ}, the green, red, and blue lines correspond to
the boundaries between the models where dust cooling, chemical heating,
and molecular cooling, respectively, are effective and not.
To draw the boundary curves, we linearly interpolate the critical $f_0$ and $Z$
obtained by the one-zone calculations.
Above the green-dotted line, dust cooling can trigger gas fragmentation (criterion {\it i}).
Below the red-dashed line, H$_2$ formation heating is not effective to suppress the
gas elongation ({\it ii}-a).
In the regions surrounded by the blue solid lines, OH and H$_2$O cooling
can enhance the cloud elongation even though the H$_2$ formation heating is effective
above the red-dashed line ({\it ii}-b).
The green-shaded region indicates the parameter region where the
criteria ({\it i}) and ({\it ii}-a) are both satisfied.
In the blue-shaded regions, the fragmentation is favored by the OH and H$_2$O cooling.
Interestingly, our criteria are satisfied in the limited metallicity 
region $Z\sim 10^{-5}$--$10^{-4} \ \Zsun$.
In the next section, we explain and derive the boundaries
using simple analytic models.

\subsubsection{Analytic formulae of the boundaries}
\label{sec:analytic}

\noindent{\bf Dust cooling}

The criterion ({\it i}) is defined by the balance between the
compressional heating rate $\Gamma _{\rm adi}$ and
the rate of heat transfer from gas to dust, $\Lambda _{\rm d}$.
In \citetalias{Chiaki15}, we define the criterion of metallicity for efficient dust
cooling as
\[
Z > 10^{-5.5} \ \Zsun 
\left( \frac{\fdepini}{0.18} \right) ^{-0.44} .
\]
The equation is defined with $f_0 = 3\pi /2 = 4.7$.
Since $\Gamma _{\rm adi} = p/t_{\rm col}$ is reciprocally proportional to $f_0$,
the more general formula of this criterion is
\[
Z > 10^{-5.5} \ \Zsun
\left( \frac{\fdepini}{0.18} \right) ^{-0.44}
\left( \frac{f_0}{4.7} \right) ^{-1} .
\]
Since it is unrealistic to measure the mass fraction of all heavy elements,
we convert this equation to
\begin{equation}
10^{\rm [Mg/H]} > 1.5f_0^{-1}\left( \frac{\fdepini}{0.18} \right) ^{-0.44}\E{-5} ,
\label{eq:fZ_Dust}
\end{equation}
using abundance of magnesium as a reference element.
In this conversion, we use the relationship $10^{\rm [Mg/H] }= Z/Z_{\bigodot}$.\footnote{Indeed, our 
Pop III nucleosynthesis models predict that 
$10^{\rm [Mg/H]} = [0.86$--$1.21] (Z/Z_{\bigodot})$ for core-collapse supernovae (CCSNe) with progenitor mass $\Mpr = 13$--$30 \ \Msun$, and 
$10^{\rm [Mg/H]} = 1.12$ and $0.84 (Z/Z_{\bigodot})$ for pair-instability supernovae (PISNe) with $\Mpr = 170$ and $200 \ \Msun$, respectively.}

\citet{Cayrel04} suggests that Mg would be more suitable as the reference
element to stellar metallicity than Ca and Fe because its main production epoch
is hydrostatic burning of progenitor stars,
and because its abundance is less affected by mixing/fallback mechanism of
SN \citep{Woosley95}.
Also, magnesium is the key element of forsterite ($\Forsterite$) grains, which is main species for
gas cooling for most Pop III SN models \citepalias{Chiaki15}, i.e.,
magnesium controls the abundance of forsterite grains because its stoichiometric number is two.
Therefore, we use magnesium as a reference species.

\vspace{5mm}
\noindent{\bf H$_2$ formation heating}

The condition where the rapid gas heating by H$_2$ formation occurs 
is that the equilibrium temperature
$T_{\rm eq}$ at $\nH <10^{8} \ \percc$ 
is smaller than that at $\nH >10^{8} \ \percc$.
At $10^{11} \ \percc$, the thermal balance between 
H$_2$ formation heating and H$_2$ line cooling yields $T_{\rm eq} \sim 870$ K.
At $\nH = 10^7 \ \percc$,
adiabatic gas compression is a major heating process, whereas the dominant
cooling mechanism is H$_2$ line transition with $Z \lesssim 10^{-5} \ \Zsun$,
or H$_2$O line transition with $Z\gtrsim 10^{-5} \ \Zsun$.
The condition where the equilibrium temperature is less than 870 K at $\nH = 10^7 \ \percc$ is
$\Gamma _{\rm adi} < \Lambda _{\rm H_2} + \Lambda _{\rm H_2O}$,
where $\Lambda _{x}$ is the cooling rate of species $x$ per unit volume.
Since oxygen is almost fully in the form of H$_2$O molecules
at this density, these cooling and heating rates are given by
\[ \Gamma _{\rm adi} = 2.8\E{-17} f_0^{-1} \ {\rm erg \ cm^{-3} \ s^{-1}}, \]
\[ \Lambda _{\rm H_2} = 3.7\E{-18} \ {\rm erg \ cm^{-3} \ s^{-1}}, \] 
\[ \Lambda _{\rm H_2O}  = 4.0\E{-18}
 \left( \frac{10^{\rm [O/H]}}{10^{-5}} \right)
 \ {\rm erg \ cm^{-3} \ s^{-1}} \]
at $10^7 \ \percc$ and $T=870$ K with $t_{\rm col}^{\rm s} (1.0) = 0.34 t_{\rm ff}$.
Thus, using oxygen abundance, we can write the condition as
\begin{equation}
10^{\rm [O/H]} > ( 6.8 f_0^{-1} - 0.91) \E{-5} .
\label{eq:fZ_H2form}
\end{equation}
The boundary is consistent with the red dashed line in Figure \ref{fig:ox_fZ}.
For a gas cloud with $f_0$ and $Z$ above both the red and green lines,
we expect that the gas does not get elongated owing to the effect of rapid H$_2$
formation heating. Hence the filament fragmentation does not occur either
when dust cooling is effective.
On the other hand,
the green shaded region above the green line and below the red line
is the regime where the dust cooling is efficient
but H$_2$ formation heating is not (criteria {\it i} and {\it ii}-a).
The cloud MH1 Z5 where the fragmentation occurs
is plotted in the green region.
We explicitly confirm the cloud fragmentation in the additional simulation 
for UNI with $10^{-4.6} \ \Zsun$.

\vspace{5mm}
\noindent{\bf OH and H$_2$O cooling}

The region surrounded by the blue solid lines indicate the regime where
OH and H$_2$O cooling is effective.
At $\nH \sim 10^7 \ \percc$, OH cooling rate exceeds the compressional heating rate.
The rates are
\[ \Gamma _{\rm adi} = 1.0\E{-17} f_0^{-1} \ {\rm erg \ cm^{-3} \ s^{-1}}, \]
\[ \Lambda _{\rm OH} = 1.2\E{-18} 
\left( \frac{10^{\rm [O/H]}}{10^{-4}}  \right) \ {\rm erg \ cm^{-3} \ s^{-1}} \]
at $10^7 \ \percc$ and 200 K with $t_{\rm col}^{\rm s} (0.5) = 0.21 t_{\rm ff}$,
where OH abundance is in general a tenth of oxygen abundance.
The condition where $\Gamma _{\rm adi} = \Lambda _{\rm OH}$ can be derived to be
\begin{equation}
10^{\rm [O/H]} = 1.6f_0^{-1}\E{-4}  .
\label{eq:fZ_OH}
\end{equation}
For large $f_0$ and $Z$, OH molecules are formed at $\nH <10^7 \ \percc$ and
$\Gamma _{\rm adi}$ balances $\Lambda _{\rm OH}$ already  at $10^7 \ \percc$.
In this case, the gas evolves nearly isothermally without significant
temperature drop.
Therefore, the condition ({\it ii}-b) is satisfied in the narrow region
around the part of the line given by Equation (\ref{eq:fZ_OH}).
The blue shaded region with the label ``OH/H$_2$O'' indicate the regimes 
where OH and H$_2$O cooling promotes the gas elongation.
For the combination of $f_0$ and $Z$ in these regions,
the gas cloud eventually fragments by dust cooling
even though H$_2$ formation heating is efficient (criteria {\it i} and {\it ii}-b).
The cloud MH1 Z4 (discussed in Section \ref{sec:OH_H2O}) is in the blue-shaded region,
where the OH cooling enhances gas elongation (Figure \ref{fig:nTe_3d}),
and gas fragmentation is finally triggered (Figure \ref{fig:snapshot_all}). 
Also for MH2 with $10^{-4.2} \ \Zsun$, the clouds are sufficiently elongated to fragment.

\begin{figure}
\includegraphics[width=8.5cm]{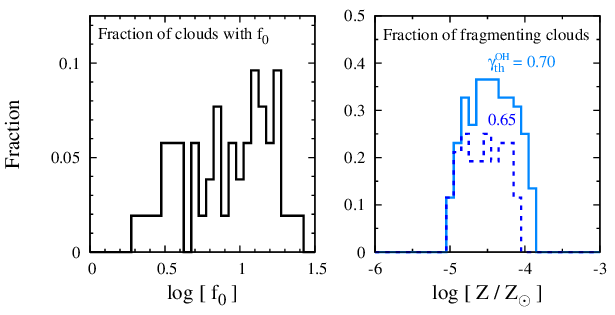}
\caption{
{\it Left panel}: number of 52 Pop III clouds with various collapse time $f_0$ simulated
by \citetalias{Hirano14}.
{\it Right panel}: number fraction of Pop II clouds which will undergo fragmentation if these clouds have the
same $f_0$ distribution as Pop III clouds.
We integrate the fraction of clouds with $f_0$ which satisfy the fragmentation condition defined by
the one-zone calculations.
The frequency depends on the threshold specific heat ratio $\gammath ^{\rm OH}$ below which cloud is considered to be
elongated sufficiently by OH cooling.
We consider two cases with $\gammath ^{\rm OH} = 0.65$ (blue) and $0.70$ (cyan).}
\label{fig:ff_Pop}
\end{figure}

\subsubsection{Fraction of low-mass star-forming clouds}
It is interesting to ask the number fraction of the clouds that yield fragments.
In order to know this, a distribution of the collapse timescale of low-metal
gas clouds is needed.
Unfortunately, we can not observationally derive the statistics of the collapse timescale of clouds.
Instead, one of the clues may be obtained from
the distribution of the collapse timescales of primordial clouds
obtained by the cosmological simulations of \citetalias{Hirano14}.
We derive $f_0$ for 52 primordial clouds 
The left panel of Figure \ref{fig:ff_Pop} shows the distribution of
$f_0$ of primordial clouds.
It spreads between $\log (f_0) = 0.3$ and $1.4$.
We then calculate the number fraction of clouds that satisfy our fragmentation criterion indicated in Figure
\ref{fig:ox_fZ} with different metallicities.
The right panel of Figure \ref{fig:ff_Pop} shows the number fraction of
Pop II clouds which will undergo fragmentation.
The fraction depends on the threshold value of specific heat ratio $\gammath ^{\rm OH}$ under which
OH cooling can enhance cloud elongation (see Section \ref{sec:analytic}).
We consider the two cases with $\gammath ^{\rm OH} = 0.65$ and $0.70$.
For severer criterion, $\gammath ^{\rm OH} = 0.65$, the fraction becomes systematically smaller.
For $10^{-5}$--$10^{-4} \ \Zsun$,
the fractions are predicted to be $\sim 25$ and $35\%$ for $\gammath ^{\rm OH} = 0.65$ and $0.70$, respectively.
It have been considered that clouds would be more likely to host low-mass stars with increasing metallicity.
However, we find that H$_2$ formation heating can prohibit fragmentation in a range of metallicity
$\gtrsim 10^{-4} \ \Zsun$.
In our model, we predict that the metallicity range between $\sim 10^{-5}$ and $\sim 10^{-4} \ \Zsun$ is favored to fragmentation.
Even in this range, clouds which undergo fragmentation is minority.
Considering that we here employ the carbon-normal elemental abundance ($\rm [C/Fe] <1$),
we argue that the number of low-mass EMP stars 
discovered so far is small \citep{Caffau11}, possibly reflecting the stringent constraints on
the fragmentation of low-metallicity gas clouds.
Here we estimate the fraction of clouds with fragmentation
with the elemental abundance of Pop III progenitor with $30 \ \Msun$.
In Section \ref{sec:discussion}, we see the metal abundances with which fragmentation
occurs for a wide range of progenitor masses.

\subsection{Disk fragmentation with $Z\lesssim 10^{-5} \ \Zsun$}
\label{sec:DiskFrag}

The fragmentation occurs also for MH1 Z6 although its $f_0$ and $Z$
are outside the region of the fragmentation condition shown in Figure \ref{fig:ox_fZ}.
That is even reasonable because the criteria should not be applied to
the cases with 
\begin{equation}
Z < 1.5f_0^{-1}\left( \frac{\fdepini}{0.18} \right) ^{-0.44}\E{-5} \ \Zsun,
\label{eq:fZ_NoDust}
\end{equation}
where dust cooling is inefficient (see Equation \ref{eq:fZ_Dust}).
Several authors have discussed that low-mass stars are likely to 
form by the fragmentation on the accretion disks even in the metal-free cases 
\citep{Clark11, Greif12, Susa14}. 
It can be considered that the fragmentation in their simulations can be 
categorized in the disk fragmentation.

In the regime, the fragmentation is not triggered by gas cooling but
would be triggered by its self-gravity \citep[e.g.][]{Vorobyov10}.
Gravitational instability of a protostellar disk has been
studied extensively \citep[e.g.][]{Gammie01}.
In our simulations, the cloud fragments only for one cloud MH1 Z6 out
of five clouds (UNI Z5 and four Z6 clouds) satisfying Equation (\ref{eq:fZ_NoDust}).
It might be because of the peculiar density distribution of the MH2 cloud
or just because the disk evolution can be followed for the longest time (16 years)
among the five clouds.
The fragmentation might be observed also for other clouds if we could follow their evolutions
for the longer time.
Here, we indicate the maximal region favorable to the disk fragmentation 
as Equation (\ref{eq:fZ_NoDust}) 
indicated by the grey-hatched area on the $f_0$-$Z$ plane of Figure \ref{fig:ox_fZ}. 

In our cloud MH1 Z6, five gas clumps are formed 
but all of them immediately migrate into the central primitive protostar
with a life time of several years.
If the formation and migration continue similarly, the mass of the central
protostar quickly increases.
If some blobs are ejected from the dense disk by multi-body dynamical interactions,
they would avoid the merger with the central star and also could no longer accrete
the gas from the disk \citep{Zhu12, Susa14}.
This would lead the formation of the low-mass stars even in the so-called
``hyper metal-poor (HMP)'' regime with metallicities $\lesssim 10^{-5} \ \Zsun$.


\begin{figure}
\includegraphics[width=8.5cm]{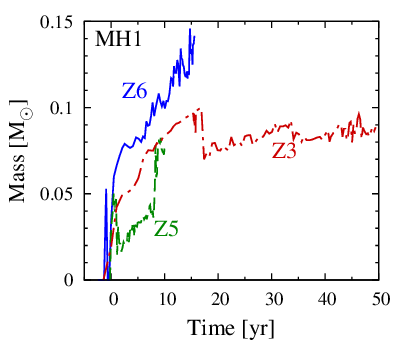}
\caption{
Mass of the protostar first formed in the cloud MH1 for
Z3 (red dot-dashed), Z5 (green dashed), and Z6 (blue solid)
as a function of the time from the formation of the first protostar.}
\label{fig:snapshot_ps_all}
\end{figure}

\begin{figure}
\includegraphics[width=8.5cm]{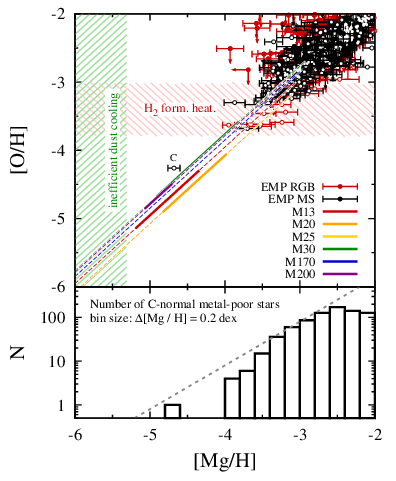}
\caption{
{\it Top panel}: magnesium and oxygen abundances
with which the relevant thermal processes affect the cloud fragmentation.
The green colored line represents the combination of Mg and O abundances predicted
by the Pop III core-collapse supernova (CCSN) model with progenitor mass $\Mpr = 30 \ \Msun$
employed in our simulations.
We also draw the abundances synthesized in CCSNe with $\Mpr = 13$, $20$, and $25 \ \Msun$
by the red, orange, and yellow lines, respectively.
The blue and purple lines are for pair-instability supernova (PISN) models respectively
with $\Mpr = 170$ and $200 \ \Msun$.
These lines are drawn up to the metallicity $Z = 10^{-3} \ \Zsun$, above which a single SN
explosion can hardly enrich clouds \citep[e.g.][]{Audouze95}.
Our one-zone model predicts that the fragmentation occurs in the solid segment for {\it existing} $f_0$, 
while not in the dashed region for {\it any} $f_0$ (with a range of $f_0=1$--$10$).
In the green-hatched region, dust cooling is ineffective because the abundance
of magnesium (key element of silicate grains) is too small (Equation \ref{eq:fZ_Dust}).
In the red-hatched region, rapid H$_2$ formation heating halts cloud elongation because temperature
declines by H$_2$O molecular cooling before the exothermic reactions of H$_2$ formation sets in
(Equation \ref{eq:fZ_H2form}).
We also plot the abundances of the extremely metal-poor (EMP) stars so far observed for comparison
taken from the SAGA database \citep[][http://sagadatabase.jp/]{Suda08}.
The open circles represent the stars whose oxygen abundance is unavailable, and predicted
from iron abundance with the average rate ${\rm [O/Fe] }= 0.47$ of the EMP stars \citep{Cayrel04}. 
The star labeled ``C'' is most metal-poor star ever observed, \Caffau \ \citep{Caffau11}.
{\it Bottom panel}: histogram of the magnesium abundances of EMP stars with bin size of
$\Delta {\rm [Mg/H]} = 0.2 \ \dex$.
The dashed black line shows the reciprocal relationship of star formation to gas metallicity
predicted by a simple cosmological enrichment model by \citep{Hartwick76}.}
\label{fig:Mg_O}
\end{figure}

\section{Discussion and Conclusion}
\label{sec:discussion}

\subsection{Summary}

The most important conclusion we draw from our set three-dimensional simulations
is that not only the gas metallicity $Z$ but also the collapse timescale $f_0$
determine the thermal evolution and fragmentation ``mode'' of
a contracting gas cloud. Thermal evolution and core fragmentation are
intrinsically connected to each other in the following manner:
\begin{itemize}
\item
The dust thermal emission induces significant deformation of a cloud core.
Because accretion of gas-phase metal onto grains enhances cooling efficiency,
it is important to follow the growth of dust grains in a low-metallicity gas.
\item
Gas heating by H$_2$ formation via three-body reactions plays a crucial role
to halt the elongation of the gas in the cases of vast majority with metallicities 
$10^{-4}$--$10^{-3} \ \Zsun$ with the early metal and dust model.
\item
OH and H$_2$O gas cooling enhances the cloud elongation and yields
favorable conditions for core fragmentation later when dust cooling
becomes efficient.
In some cases, this occurs before H$_2$ formation heating becomes efficient.
\end{itemize}
We can qualitatively explain the conditions of filamentary
fragmentation on a $f_0$-$Z$ plane as shown in Figure \ref{fig:ox_fZ}.

\subsection{Origin of $f_0$}
The difference of cloud collapse time would result from
the depth of potential well created by hosting dark matter halos.
We find that the depth $G \Mvir / \Rvir$ (Table \ref{tab:cloud}) negatively
correlates with the collapsing time $f_0$ for the 52 primordial clouds including MH1,
MH2, and MH3 simulated by \citetalias{Hirano14}.
This shows that, as the potential well becomes deeper, the collapse
time becomes shorter.
We also find that gas rotation hardly affects the collapse timescale.

\subsection{Comparison with the observations}
In our simulation, we employ the progenitor model only
with mass $\Mpr = 30 \ \Msun$ and ambient gas density $\namb = 1 \ \percc$.
In this progenitor model, we have seen that oxygen and magnesium can control the cooling/heating rates important 
for the cloud fragmentation (Section \ref{sec:analytic}).
The combination of Mg and O abundances are drawn by the green line in Figure \ref{fig:Mg_O}.
Our one-zone calculations show that $f_0$ for which the cloud would fragment {\it exists} with metallicities 
$-5.0 < \log (Z/Z_{\bigodot} ) < -3.9$ (Figure \ref{fig:ox_fZ}).
This region is shown by the solid thick segment in Figure \ref{fig:Mg_O}.
Similarly, the region of Mg and O abundances for other core-collapse supernova (CCSN) 
models with $\Mpr = 13$, $20$, and $25 \ \Msun$
and pair-instability supernova (PISN) models with $\Mpr = 170$ and $200 \ \Msun$
is shown by red, yellow, orange, blue, and purple lines, respectively, in Figure \ref{fig:Mg_O}.
These lines are drawn up to the corresponding metallicity $Z=10^{-3} \ \Zsun$, above which a
single SN can hardly enrich a cloud, i.e., our assumption that the cloud is polluted by
a single SN would fail \citep{Audouze95}.

For progenitor masses $\Mpr = 20$ (orange) and $200 \ \Msun$ (purple), 
OH/H$_2$O cooling operates as in the case with $\Mpr = 30 \ \Msun$.
For $\Mpr = 13 \ \Msun$, the molecular cooling can not affect the cloud fragmentation because
oxygen is less abundant ($ {\rm C/O}>1$).
Instead, carbon grains can grow and contribute to fragmentation.
For $\Mpr = 25 \ \Msun$ (yellow), since ${\rm C/O} < 1$, carbon grains can not grow to enhance
cloud fragmentation.
Nevertheless, since the ratio ${\rm C/O}$ is still less than $\Mpr = 20$ and $30 \ \Msun$ models \citep[also see][]{Nozawa03},
cooling and silicate grains is not important.
In PISN case with $\Mpr = 170 \ \Msun$ (blue), silicate grains have already been significantly destroyed by the
reverse shock just after the explosion.
In the models with $\Mpr = 25$ and $170 \ \Msun$, even the criterion ({\it i}) is not satisfied for any $f_0 = 1$--$10$.

In the most cases of carbon-normal star formation, we could say that
Mg and O abundances are important to predict the fragmentation
property of the prestellar clouds.
Instead, the fragmentation of carbon-enhanced cloud should controlled by other factors
such as carbon abundance \citep[see][]{Marassi14}.
Even in our progenitor model with $\Mpr = 13 \ \Msun$, we see that the C {\sc i} cooling
creates the steep temperature decline.
Whether this decline can enhance the cloud elongation or not can not be determined by
the one-zone model.
We will discuss this in the forthcoming paper.

The one-zone model predict the region where cloud fragmentation
is disfavored because dust cooling is insufficient by small
amount of magnesium is shown by the green-hatched region in Figure \ref{fig:Mg_O}.
In the red-hatched region, cloud elongation is prevented by H$_2$ formation heating.
With metallicities $Z\gtrsim 10^{-3} \ \Zsun$, the abundance ratio approaches
the present-day value ($\rm [O/H] = [Mg/H]$), and the dust properties would also
approach the one in the local interstellar medium.
If so, we can predict that H$_2$ formation heating would prevent fragmentation in the range $\rm -4 \lesssim [O/H] \lesssim -3$.

Figure \ref{fig:Mg_O} also shows the magnesium and oxygen abundances of EMP stars so far observed
by filled circles.
Unfortunately, oxygen abundance of considerable stars is unavailable because
the oxygen line is generally under the detection limit or blended with the telluric lines \citep{Cayrel04}.
To secure the statistical samples, we resort to iron abundance of some EMP stars
whose oxygen abundance is not detected.
We convert the iron abundance to oxygen abundance, using the ``typical'' ratio ${\rm [O/Fe]} = 0.47$
\citep{Cayrel04}.\footnote{The rate {\rm [O/Fe]=0.47} is obtained from 3D stellar atmosphere model.
The 1D model predicts ${\rm [O/Fe]} = 0.7$.}
These abundances are shown by open circles in Figure \ref{fig:Mg_O}.
In the bottom panel of Figure \ref{fig:Mg_O}, we show the frequency
of the magnesium abundance of EMP stars.

The observations have so far found no EMP stars in the green region
where dust cooling fails.
Also, the less number of stars are found in the red-hatched region where H$_2$
formation heating is promoted.
The simple enrichment model of \citet{Hartwick76} suggests that the star formation
rate is reciprocally proportional to the metallicity.
The departure of the observed frequency (black boxes in Figure \ref{fig:ox_fZ})
from the predicted reciprocal relationship (black dashed line)
below the metallicity $\sim 10^{-3.5} \ \Zsun$ might be owing to the
effect of H$_2$ formation heating \citep[also see][]{TsuribeOmukai08}.
Also, the abundance of the star \Caffau \ (labeled by ``C'') is not inconsistent with the
isolated region favorable to cloud fragmentation drawn by solid lines.
Both further model predictions and observations should be required to 
give a determinative conclusion for the origin of the EMP and low-mass stars in the Galactic halo.

\subsection{Other important physics}
It is important to notice that the formation of low-mass fragments
does not immediately suggest the formation of low-mass, low-metallicity
stars. 
The final mass of a star is determined when its protostellar evolution reaches
the zero-age main sequence, i.e., after the whole history of gas accretion from the circumstellar disk
starting from a tiny embryo through growth and contraction
\citep{Shu77,Shu87,Omukai03,Hosokawa11,Hosokawa12}.
Since the dynamical time becomes progressively small with increasing density
as $\propto \rho ^{-1/2}$, we are able to follow the gas accretion process only
several decades after the first protostar formation because of technical difficulties.
There are a number of important physical processes that affects the growth
of a protostar after the initial phase.
For example, when the first, central protostar grows to $\sim 10 \ \Msun$,
photodissociation and photoionization by the ultraviolet emission may become effective to suppress the gas accretion
onto the central star \citep{Hosokawa12}.
The magnetic field would also affect the accretion process \citep{MachidaDoi13}.

We follow the gas accretion onto protostars for a limited time (a few tens
of years) after the first protostar is formed 
because we do not employ the sink-particle technique.
However, we can accurately follow the process of the gas accretion including
the merger of protostellar cores.
We plot the accretion history of the central protostar for MH1 in Figure
\ref{fig:snapshot_ps_all}.
As for the other clouds, we find that the protostar accretes a larger amount
of gas with lower metallicities.
We also comment that 
the mass accretion rate onto the protostar for MH1 is smaller than for UNI and MH2
even with the same metallicity.
This is because the gas temperature remains lower in a slowly contracting gas cloud.
Furthermore, the accreting gas is shared by several fragments,
and the growth rate of each core is correspondingly reduced.
One can naively guess that the fraction of low-mass stars would increase with
increasing metallicity.

In a realistic case where the gas is directly enriched by Pop III SNe,
significant turbulence would be generated.
Turbulence can suppress the growth of density perturbations,
or rather can even promote gas fragmentation 
\citep{Dopcke11,Dopcke13,Smith15}.
\citet{Dopcke13} report that small mass fragments (sink-particles)
are formed in turbulent gas 
for a wide range of metallicities from 0 to $10^{-3} \ \Zsun$.
They also find that the mass function of sink particles transforms from
flat one to bottom-heavy with increasing metallicites
between $10^{-5}$ and $10^{-4} \ \Zsun$.
\citet{Smith15} find vigorous fragmentation of a gas that is
enriched with a metallicity $2\E{-5} \ \Zsun$
by the Pop III SN exploding on a neighboring halo.
Their result supports the notion that turbulence-driven disturbances in a
low-metallicity gas induce gas fragmentation.
It is still uncertain how the sink particle technique employed in their simulations 
would affect the fragmentation properties.
Further studies are certainly needed to 
examine the effects of turbulence on gas fragmentation
in a fully cosmological set up.

\section*{acknowledgments}

We thank K. Omukai for fruitful discussion.
The Pop III dust model is calculated by T. Nozawa.
We could make our manuscript highly improved by the anonymous
referee's comments.
GC and SH are supported by Research Fellowships of the Japan
Society for the Promotion of Science (JSPS) for Young Scientists.
NY acknowledges the financial supports from JST CREST
and from JSPS Grant-in-Aid for Scientific Research (25287050).
The numerical simulations are carried out on Cray XC30 at Center for Computational 
Astrophysics, National Astronomical Observatory of Japan and 
on COMA at Center for Computational Sciences in University of Tsukuba.



\end{document}